\documentclass[10pt,preprint]{aastex}
\slugcomment{Version: \today}

\usepackage{color}
\usepackage{graphicx}
\usepackage{amsmath}

\def \simless {\mathbin{\lower 3pt\hbox{$\rlap{\raise 4pt
              \hbox{$\char'074$}}\mathchar"7218$}}}
\def \simgreat {\mathbin{\lower 3pt\hbox{$\rlap{\raise 4pt
              \hbox{$\char'076$}}\mathchar"7218$}}}
\def\ie {{\it i.e.}}
\def\cf {{\it c.f.}}

\begin{document}
\title{Steady Wind-Blown Cavities within Infalling Rotating Envelopes: Application to the Broad Velocity Component in Young Protostars 
}
\author{
	Lichen Liang\altaffilmark{1},
        Doug Johnstone\altaffilmark{2,3},
        Sylvie Cabrit\altaffilmark{4},        
        \& Lars E. Kristensen\altaffilmark{5}
}
\altaffiltext{1}{Institute for Computational Science, University of Zurich, Zurich CH-8057, Switzerland}
\altaffiltext{2}{National Research Council Canada, Herzberg Astronomy \& Astrophysics, 5071 West Saanich Rd, Victoria, BC, V9E
2E7, Canada; doug.johnstone@nrc-cnrc.gc.ca}
\altaffiltext{3}{Department of Physics \& Astronomy, University of Victoria, Victoria, BC, V8P 5C2, Canada}
\altaffiltext{4}{Observatoire de Paris, PSL University, Sorbonne Universit\'{e}, CNRS, LERMA, 61 Av. de l'Observatoire, 75014, Paris, France}
\altaffiltext{5}{Niels Bohr Institute \& Centre for Star and Planet Formation, University of Copenhagen, \O ster Voldgade 5--7, 1350 Copenhagen K,
Denmark}

\begin{abstract}
Wind-driven outflows are observed around a broad range of accreting objects throughout the Universe, ranging from forming low-mass stars to  super-massive black holes. 
We study the interaction between a central isotropic wind and an infalling, rotating, envelope, determining the steady-state cavity shape formed at their interface under the assumption of weak mixing. 
The shape of the resulting wind-blown cavity is elongated and self-similar, with a physical size determined by the ratio between wind ram pressure and envelope thermal pressure. 
We compute the growth of a warm turbulent mixing-layer between the shocked wind and the deflected envelope, and calculate the resultant broad line profile, under the assumption of a linear (Couette-type) velocity profile across the layer.
We then test our model
against the warm broad velocity component observed in CO $J$=16--15 by \textit{Herschel}/HIFI in the protostar Serpens-Main SMM1. Given independent observational constraints on the temperature and density of the dust envelope around SMM1, 
we find an excellent match to all its observed properties  
(line profile, momentum, temperature) and to the SMM1 outflow cavity width
for a physically reasonable set of parameters: a ratio of wind to infall mass-flux $\simeq 4\%$,
a wind speed $v_{\rm w} \simeq 30$ km/s, 
an interstellar abundance of CO and H$_2$,
and a turbulent entrainment efficiency consistent with laboratory experiments. 
The inferred ratio of ejection to disk accretion rate, $\simeq 6-20\%$, 
is in agreement with current disk wind theories. Thus, the model provides a new framework 
to reconcile the modest outflow cavity widths in protostars with the large observed flow velocities. 
Being self-similar, it is applicable over a broader range of astrophysical contexts as well.
\end{abstract}


\section{Introduction}
\label{sec:intro}

Massive outflows are observed everywhere in the Universe, ranging from individual forming stars through galactic-scale events. When supersonic stellar or galactic winds interact with the surrounding medium, be it the molecular envelope around forming stars or the intergalactic gas, they are observed to impart momentum and energy and entrain a slower-moving massive outflow. The actual entrainment mechanism and efficiency, however, remain poorly understood and highly debated, both because of a lack of strong observational constraints as well as a  relative paucity of theoretical predictions against which to test observations. 

A specific example of entrainment takes place when an accreting protostar launches a highly collimated jet, possibly surrounded by a wider-angle disk wind, carving out a large and slow massive outflow cavity into the parent cloud \citep{frank14}. When the \textit{Herschel} Space Observatory \citep{pilbratt10} started observing protostars in H$_2$O and high-$J$ CO rotational transitions, it quickly became clear that the dominant source of the emission was from molecular outflows \citep[e.g.,][]{vandishoeck11, Kristensen12, Kristensen17}. It also became clear that these emission lines highlight a different outflow component from the low-$J$ CO transitions observed from the ground, such as $J$=2--1 and 3--2 \citep[e.g.,][]{yildiz13}. This \textit{Herschel}-bright outflow component has both a significantly higher temperature $\simeq 200-500$ K and a larger line width at half maximum (FWHM) $\ge 30$ km~s$^{-1}$ compared to low-$J$ CO line profiles, where it only appears as a faint pedestal in very deep integrations \citep{margulis89}.
Accordingly, it was labeled the ``broad'' outflow component \citep{Kristensen12, Kristensen17, mottram14, mottram17}. 


The physical origin of the ``broad" warm outflow component, and its relation to both the slower cold outflow, seen in low-$J$ emission, and the faster protostellar jet or wind is not clear. Two hypotheses have been put forward: either this broad component arises within a warm and dusty disk wind \citep{panoglou12, yvart16} or it arises where ambient material is currently being entrained into the outflow by the protostellar wind, for example through non-dissociative shock waves \citep{Kristensen12, Kristensen17, mottram14, mottram17}. While detailed, dynamical, and thermo-chemical predictions exist for the disk wind models, which reproduce the observed H$_2$O emission \citep{yvart16}, only limited model predictions exist in the literature for the entrainment scenario, and it thus remains a hypothesis. The underlying physical issue is not a problem reserved for protostellar outflows, but remains an uncertainty for outflows in general. 

A first type of entrainment scenario proposes that outflows are entrained by large jet bowshocks. These models predict substantial warm molecular material at intermediate velocities \citep[e.g.][]{rc93, downes03},
but the resulting outflow cavities have been deemed too elongated compared to observations \citep{ostriker01}. To avoid this potential issue, a second type of entrainment scenario proposes that entrainment is dominated instead by a wide-angle wind. A particularly popular version of this ``wind-driven" scenario assumes instantaneous full mixing between the shocked isotropic wind and the shocked envelope material. Due to the complete mixing, the cavity retains  a primarily radial outflow motion with a roughly constant expansion speed over time \citep{li96,lee00}, except very close to the disk mid-plane where the cavity remains trapped near the  outer disk radius \citep{Wilkin03,MCR04,Lopez19}. Such a radially expanding wind-blown cavity, however, grows too quickly: with expansion speeds $\ge 10$ km/s similar to those observed in the broad component, it exceeds the typical radius $\leq 3000$ au of protostellar outflow cavities \citep[see e.g.][for HH212 and L1157, respectively]{lee15,gueth98} in only a few 1000 yrs \citep{shang06, Lopez19}. This timescale is much shorter than the typical age accepted for Class 0 outflow sources \citep[$10^4-10^5$ yrs, e.g.,][and references therein]{Kristensen18}. 

In order to circumvent this ``age" problem, in this paper we consider wind-driven cavities with partial mixing, instead of full mixing, and explore \textit{stationary solutions} for the cavity shape, formed as the fast stellar wind is obliquely deflected by the envelope and forced to flow \textit{along} the cavity wall, instead of radially outwards. Indeed, numerical simulations of a spherical wind propagating into a rotating and infalling slab show that when mixing is inefficient, the cavity ``flanks" quickly converge to a quasi-steady shape \citep{Delamarter00}. A modest outflow cavity width can then be maintained over the whole duration of the Class 0 phase. Such a configuration is also prone to the development of a turbulent mixing-layer at the contact discontinuity between the shocked wind and the shocked envelope gas, as they slide past one other \citep{rcc95}. In this paper we will therefore follow a deliberate path to testing whether such a mixing-layer might explain the broad spectral component observed around protostars by \textit{Herschel}/HIFI, and at the same time the observed cavity sizes.

Steady wind cavity solutions were first computed by \citet{BC81} for an isotropic wind expanding into a thick isothermal self-gravitating toroid. \citet{smith86} showed that similarly elongated ``flame-like" cavities could also be obtained for isotropic envelopes with a purely radial  pressure profile $p(r) \propto r^{-n}$, provided that $n < 2$ and the wind is obliquely deflected at its closest point of impact (e.g.\ by a small-scale thin disk). Both of these early calculations show that the addition of a dense, disk-like, component along the horizontal axis can provide the required equatorial pinch to create steady, elongated outflow shapes similar to those observed around protostars. Similar physical conditions are expected for galactic-scale outflows \citep[see e.g.][]{aalto16}. In this paper, we proceed one step further than these previous investigations by computing stationary solutions in a more realistic density and velocity distribution for the envelope, namely the more sophisticated \citet{ulrich76} infalling and rotating solution. Despite an identical ambient density distribution, our cavity morphologies will strongly differ from the calculations of \citet{Wilkin03}, \citet{MCR04}, and \citet{Lopez19} in that we assume {weak} mixing, instead of full-mixing, between the shocked wind and the shocked envelope, and we include the effect of thermal pressure in the envelope. These two ingredients allow the existence of stable stationary solutions on large scales, with pointed shapes at the pole.

In Section \ref{sec:shape} we determine the stationary cavity shape formed by a wide-angle wind deflected by an infalling and rotating protostellar envelope.
In Section \ref{sec:flow} we consider the deflected wind material flowing along the cavity boundary,  and the turbulent entrainment of envelope material within a mixing-layer. We then, in Section \ref{sec:profile}, compute the angular momentum and synthetic line profiles associated with material within the mixing-layer. Next, we quantitatively compare the  model results against observations from \textit{Herschel} of the broad CO component in the protostar Serpens-Main SMM1 (Section \ref{sec:AppC0}. Finally, in Section \ref{sec:conc} we conclude with a recap of the implications of these results.

\section{Determination of the Cavity Shape}
\label{sec:shape}

In this section we 
produced by an isothermal \citet{ulrich76} infalling and rotating envelope model 
interacting with an isotropic wind.
We determine the fundamental non-dimensional parameters and characteristic values 
and perform a stationary solution analysis in order to determine the range of cavity shapes produced.
The shape of the thin shell formed by this interaction is determined by the (ram plus thermal) pressure balance between the wind and the envelope along with a ``centrifugal term" due to the upward curving motion of the shocked wind layer. 

The analysis in this section,  as well as Section \ref{sec:flow} and Section \ref{sec:profile}, is presented in dimensionless form in order to focus on the underlying self-similarity of the shape and the scaling relations underpinning the solutions, which have a general applicability for all manner of outflows. In Section \ref{sec:AppC0}, 
we adopt appropriate physical values in order to quantitatively determine the agreement between the model and observations, in the specific case of the warm CO outflow of a nearby protostar.

\subsection{Infalling and Rotating Envelope Model}
\label{sec:shape:infall}

The \citet{ulrich76} infalling envelope model generalizes from the case of an isothermal cloud by combining the spherical infall of envelope material under the force of gravity due to the mass of the central object $M_\star$ \citep{bondi52} together with a treatment of the centrifugal deflection of the flow due to initial solid body rotation. Thus, in the outer envelope, before angular momentum becomes dominant, the density structure is almost spherical with $\rho(r) \propto r^{-3/2}$. At smaller radii, where centrifugal forces dominate, the flow streamlines are deflected toward the mid-plane, creating a disk-like structure inside of the fiducial radius
\begin{equation}
r_{\rm d} = \frac{\Gamma_\infty^2}{G\,M_\star}.
\end{equation}
Here, $\Gamma_\infty$ is the specific angular momentum in the equatorial plane.
Assuming a ballistic solution for the infalling material, the entire envelope solution can be described by a handful of parameters: the mass of the protostar, $M_\star$, the size of the disk, $r_{\rm d}$, and the mass infall rate, $\dot M_{\rm inf}$. 

Following \citet{ulrich76} we therefore find for the density in the envelope
\begin{equation}
  \rho_{\rm inf} (r,\theta)= \frac{\dot{M}_{\rm inf}}{8\pi(G M_* r^3)^{1/2}} \Bigg(1+\frac{\sin\theta}{\sin\theta_0} \Bigg)^{-1/2} \Bigg(\frac{\sin\theta}{{\rm 2\,sin}\theta_0} +\frac{r_{\rm d}\, \sin^2{\theta_0}}{r}\Bigg)^{-1},
\label{E5}
\end{equation}
where $r$ is the spherical radius, $\theta$ is measured from the disk plane, and the subscript $``0"$ denotes the initial value at very large distance from the origin. At any location $(r, \theta)$ in the infalling envelope, the initial angular origin of the streamline, $\theta_0$, can be obtained by solving
\begin{equation}
r=\frac{r_{\rm d}\;\sin\theta_0\;\cos^2\theta_0}{\sin\theta_0-\sin\theta}.
\label{E6}
\end{equation}

Finally, the three components of the velocity of the infalling material at position $(r, \theta)$ are given by
\begin{equation}
v_{\rm inf, r} = -\Bigg(\frac{GM_*}{r}\Bigg)^{1/2}\Bigg(1+\frac{\sin\theta}{\sin\theta_0}\Bigg)^{1/2},
\label{E7}
\end{equation}
\begin{equation}
v_{\rm inf, \theta} = -\Bigg(\frac{GM_*}{r}\Bigg)^{1/2}(\sin\theta_0-\sin\theta)\Bigg(\frac{\sin\theta_0+\sin\theta}{\sin\theta_0\cos^2\theta}\Bigg)^{1/2},
\label{E8}
\end{equation}
and
\begin{equation}
v_{\rm inf, \phi} = -\Bigg(\frac{GM_*}{r}\Bigg)^{1/2}\frac{\cos\theta_0}{\cos\theta}\Bigg(1-\frac{\sin\theta}{\sin\theta_0}\Bigg)^{1/2}. 
\label{E9}
\end{equation}
Note, we have corrected the typographical error in equation 8 of \citet{ulrich76} as noted by \citet{tobin12}.

\subsection{Wide-angle Wind}
\label{sec:shape:wind}

In this paper, we assume a spherical isotropic wide-angle wind of constant speed $v_{\rm w}$ and mass-loss rate $\dot M_{\rm w}$, with the  density profile
\begin{equation}
  \rho_{\rm w} =  \frac{\dot{M}_{\rm w}}{4 \pi v_{\rm w} r^2}.
\label{E10}
\end{equation}
This approach enables a useful comparison with previous work \citep{BC81, smith86, Wilkin03,MCR04,Lopez19} and an applicability to a wide range of astrophysical contexts. 

For example, while observations of young stars and Class 0 protostars show a strong and fast jet-like component along the axis, surrounded by a (seemingly) mostly empty lower-velocity outflow cavity, several MHD models predict that the jet may only be an ``optical illusion" and may represent only the central densest core of a wider-angle wind, launched either from the inner disk edge
\citep[``X-wind" model,][]{shang98}
 or from a larger portion of the disk surface \citep[``D-wind" model,][]{cabrit99}. 

Thus, in the context of protostellar outflows, the isotropic wide-angle wind provides an acceptable approximation to the X-wind in equatorial regions, where the interaction is the most critical to define the overall cavity shape (see below). We note that the addition of a strongly directed jet-like wind enhances breakout along the outflow axis and will thus modify the cavity shape in the polar regions. 
Comparison with observations should thus focus on regions close to the flow base, at wide angles to the flow axis.

\subsection{Determining Fundamental Non-Dimensional Parameters and Characteristic Values}
\label{sec:shape:values}

 The trapping, breakout, and early evolution of the cavity formed by an isotropic wind colliding against an \citet{ulrich76} infalling envelope was first calculated under the full-mixing hypothesis, including stellar gravity and various degrees of wind collimation \citep{Wilkin03}, and envelope rotation \citep{Lopez19}. Full-mixing requires that the shell expansion is almost radial, hence the effect of thermal pressure in the envelope was neglected compared against the infall ram pressure in the frame of the expanding shell. A simplified study of the asymptotic shell expansion based on simple ram pressure balance, was conducted by \citet{MCR04}. The authors show that it depends only on a single free parameter, namely the ratio of wind ram pressure to the fiducial infall ram pressure at $r_d$,
\begin{equation}
    \lambda \equiv \frac{ v_{\rm w}\, \dot{M}_{\rm w}}{v_{\rm d} \, \dot{M}_{\rm inf}},
    \label{eqn:lambda}
\end{equation}
where 
\begin{equation}
v_{\rm d} = (G\,M_*/r_{\rm d})^{1/2}
\end{equation}
is the Keplerian velocity at $r_{\rm d}$. 

Trapped solutions with sizes less than $r_{\rm d}$ were found for $\lambda \simless 1/2$ \citep[see Figures 5 and 8 from][]{MCR04}.
For values of $\lambda > 1/2$, the cavity solutions were found to break out and expand forever, remaining pinched only along the disk mid-plane near $r_{\rm d}$. Similar results were found by \citet{Wilkin03,Lopez19}, with an additional breakout criterion on $v_w/v_d$ for the polar cap to escape stellar gravity. 
For the X-wind and D-wind models currently favored in protostars, the denser and faster jet-like components along the axis will greatly facilitate breakout along the pole compared with the requirements for an isotropic wind \citep[cf. discussion in][]{Wilkin03}. Therefore, here we will assume that initial breakout has occurred and not consider this velocity constraint in our models.

In the present investigation, we are interested in finding steady asymptotic solutions to these breakout scenarios. For this, we assume instead that at most weak mixing occurs between the shocked wind and the shocked envelope material. Under this assumption, these two shocked and deflected layers will flow past each other, with an intermediate thin mixing-layer developing along the contact discontinuity between them. 

A further difference between our models and those of \citet{Wilkin03, MCR04, Lopez19} is that 
we take into account the role of thermal pressure in the envelope in confining the shell.  This is the crucial element allowing to reach a steady configuration on large scales, instead of infinite expansion.
Given that the density distribution in the envelope retains a modified, $r^{-3/2}$ power-law,
we anticipate that the resulting steady cavities will appear similar to the \citet{smith86} elongated outflow cavities. 

We thus introduce a characteristic scale length, $r_{\rm s}$, as the location where the ram pressure in the wind, $\rho_{\rm w}\,v_{\rm w}^2$, is balanced by the thermal pressure in the equivalent spherically symmetric infalling envelope, $\rho_{\rm inf}\,c_{s}^2$. Solving this equality yields:
\begin{equation}
{r_{\rm s}} = \left(\frac{v_{\rm w}\,\dot{M}_{\rm w}}{\dot{M}_{\rm inf}}\right)^2 
\frac{2GM_\star}{c_s^4} = \frac{2GM_\star v_0^2}{c_s^4}
\label{eqn:rs}
\end{equation}
where we define the useful characteristic velocity $v_0$, fixed by the source ejection versus accretion physics, as
\begin{equation}
v_0 \equiv v_{\rm w}\, \left(\frac{\dot{M}_{\rm w}}{\dot{M}_{\rm inf}}\right) = {c_s^2}\,\sqrt{\frac{r_{\rm s}}{2GM_\star}}.
\label{eqn:v0}
\end{equation}

Another region where thermal pressure will dominate infall ram pressure  is near the equator, where the cavity is strongly pinched at $r \simeq r_d$ \citep[see][]{Lopez19} such that infall motions become almost parallel to the cavity walls. This introduces a second fundamental non-dimensional parameter in our model, $\Lambda$, which is proportional to the ratio of wind ram pressure to envelope thermal pressure at $r_d$:
\begin{equation}
\begin{split}
\Lambda &\equiv \sqrt{\frac {2r_{\rm s}}{r_{\rm d}},}\\
    &= 2 \lambda \mathcal{M}_{\rm d}^2,
\end{split}
\label{eqn:Lambda}
\end{equation}
where 
\begin{equation}
    \mathcal{M}_{\rm d} = \frac{v_{\rm d}}{c_s}
\end{equation}
is the Mach number of the infall velocity at $r_{\rm d}$. We show in the next section that $\Lambda$ only affects the cavity shape very close to the disk mid-plane, where it 
determines the initial foot-point and opening angle.

While the disk centrifugal radius $r_{\rm d}$ provides an appropriate scaling for the geometry at the flow base, we expect $r_{\rm s}$ to provide an appropriate scaling for the geometry at large distances from the disk, where the cavity is confined by the thermal pressure in the envelope, rather than by the infall ram pressure. 


Finally, we note that while trapped solutions confined by gravity on small scales, $\le r_d$, were shown to be unstable, 
our steady shells on large scales, $\gg r_d$, are confined by thermal and ram envelope pressure and thus expected to be stable \citep[see discussion in][]{Wilkin03}. Moreover, our assumption of weak (instead of full) mixing allows a non-radial escape route for the bulk of the material reaching the shell, by either moving upward (wind) or downward (envelope).
Such situations are much more stable against wind or infall variations than the full mixing case. This is
supported by the robustness of the shell shape with respect to changes in initial or global parameters.

\subsection{Calculating the Cavity Shape}
\label{sec:shape:shape}

To determine the location of the static boundary where the wind interacts obliquely with the infalling envelope, we follow the formalism of \citet{MJH09} (their equations [2-5], which are derived in the appendix to that paper) which keeps track of both the mass and momentum flux deposited along the boundary by the shocked wind on the inner side and by the infalling material on the outer side. 

The starting condition that $\partial r/\partial \theta = 0$ at the pole, used to compute cavity shapes with full mixing \citep[eg.][]{Lopez19}, is no longer a requirement in the case of weak mixing, where flame-like shapes are allowed. Instead, we integrate from the disk mid-plane up, following \cite{smith86}. A starting location in the disk, which also explicitly sets the angle of incidence, is thus required in order to solve these equations. As noted above, \citet{Lopez19} find that breakout solutions lead to a strong pinch on the disk-plane near $r \lesssim r_{\rm d}$. We have performed a detailed analysis of the possible angles of incidence allowed at the mid-plane as a function of $r/r_{\rm d} < 1$ for all combinations of $\lambda$ and $\Lambda$ (see Appendix \ref{ap:base}). We find that when the cavity is forced to meet the mid-plane at $r/r_{\rm d} \ll 1$, the required angle of incidence at the mid-plane is such that the infalling streamlines approach the cavity wall from inside the cavity - an unphysical solution. As the mid-plane crossing approaches $r_{\rm d}$, however, there always exists a location where the angle of incidence of the cavity with the mid-plane is parallel to the infalling envelope streamlines.  We therefore use this location as our foot-point for the cavity wall. Solutions close to this starting position quickly converge above the disk to the same surface; therefore, the exactness of the starting position is not critical for these models.

\begin{figure}[ht]
    \begin{center}
    \includegraphics[width=0.4 \columnwidth]{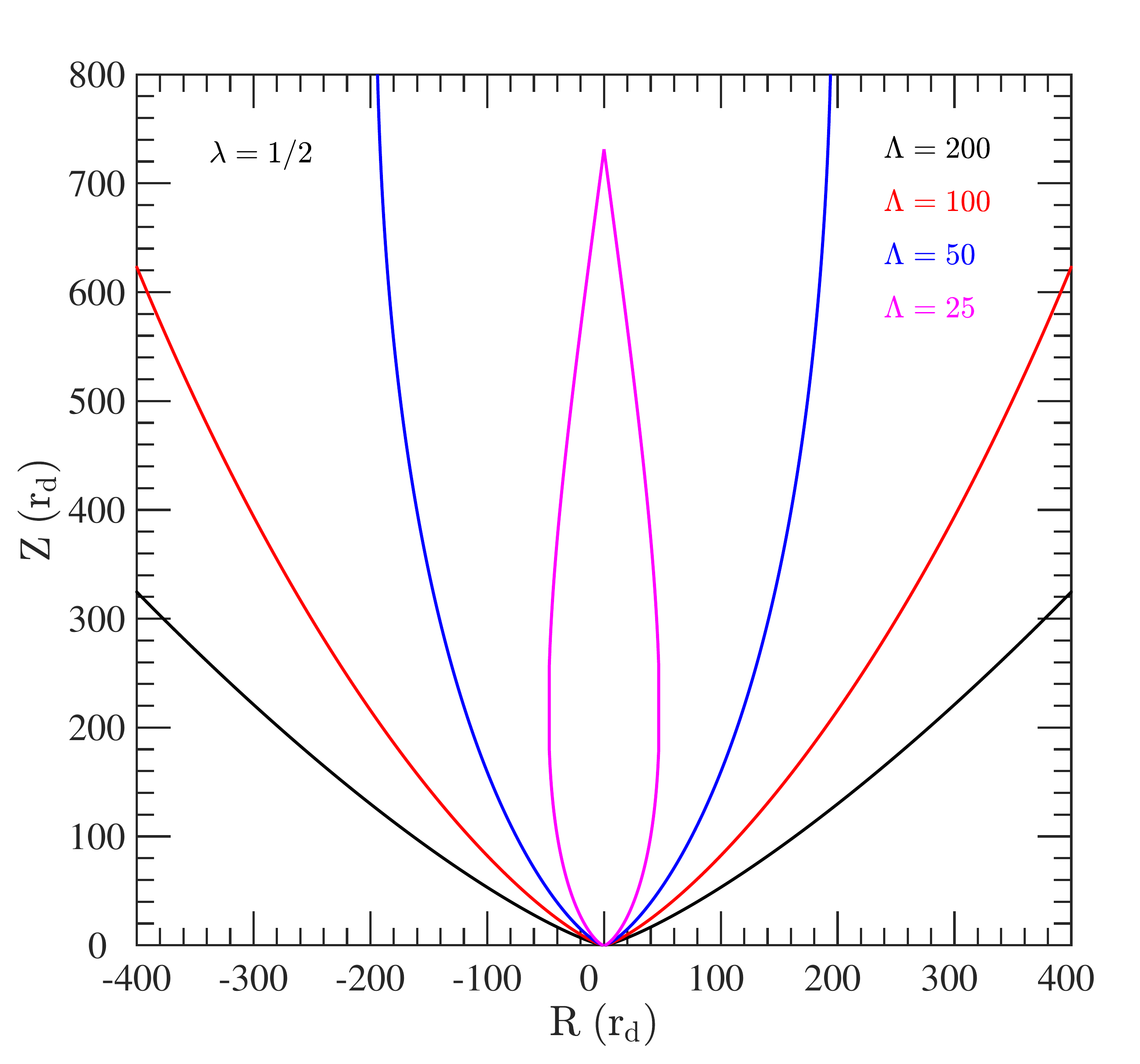}
        \includegraphics[width=0.4 \columnwidth]{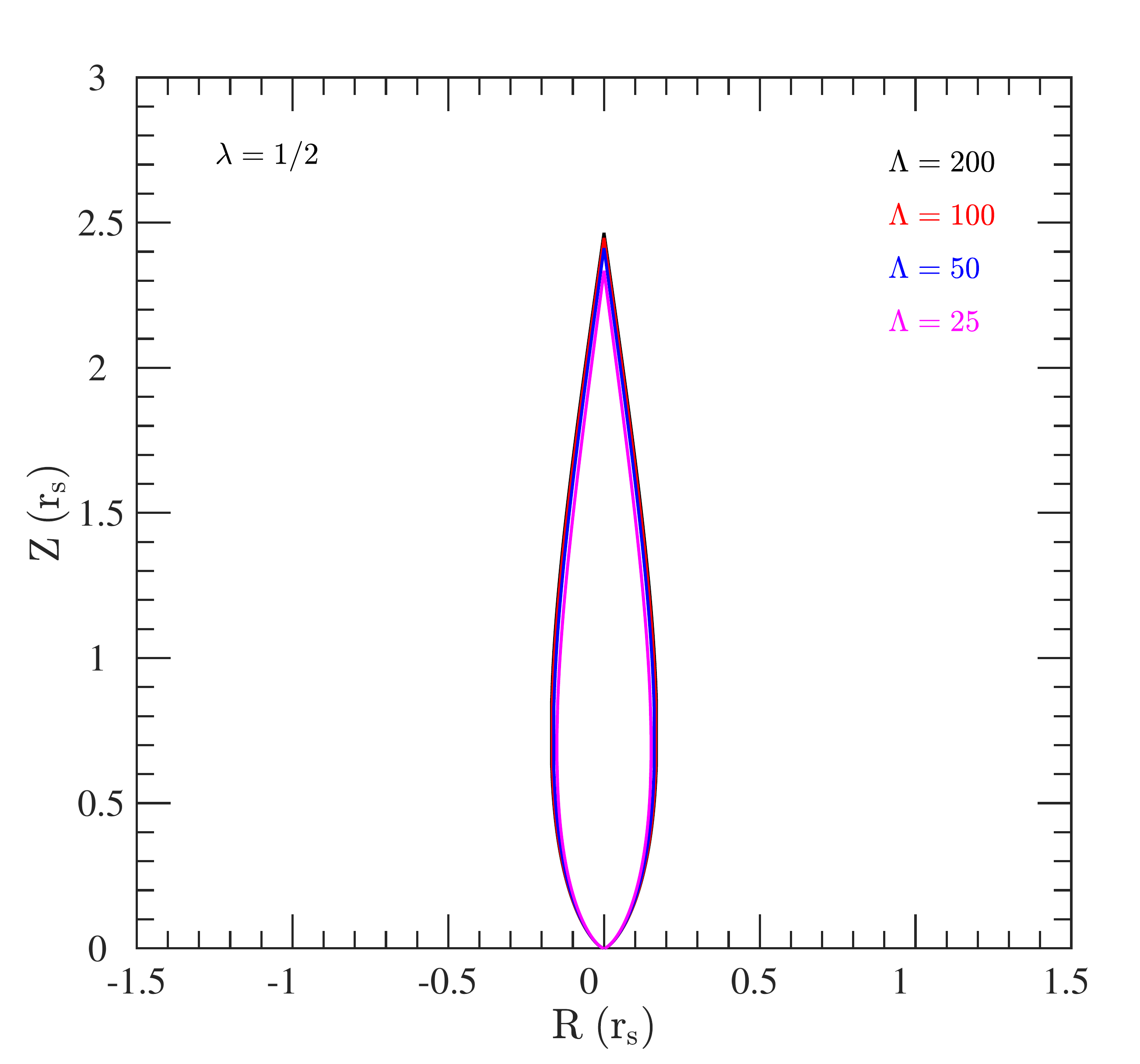}
	\end{center}
    \caption{
    Cavity shapes for an isotropic wind colliding with an \citet{ulrich76} infalling envelope for a range of $\Lambda$ values (in all cases $\lambda = 1/2$). The left panel plots the results in units of $r_d$. The right panel reveals the self-similarity of the solutions under the transformation to $r_{\rm s} = 0.5\,\Lambda^2\,r_{\rm d}$  $\equiv 2\,GM_\star (v_{\rm w} \dot M_w/\dot M_{\rm inf})^2 c_s^{-4}$, which determines the physical scale of the cavity.
     }
    \label{fig:shape}
\end{figure}

The left panel of Figure \ref{fig:shape} plots the shape of the cavity scaled by  $r_{\rm d}$, for a variety of values of $\Lambda$, while fixing $\lambda = 1/2$ (therefore $\Lambda = \mathcal{M}_{\rm d}^2$). With this scaling, the envelope appears broadest and tallest when $\Lambda$ is large due to the larger ratio of ram pressure in the wind to thermal pressure in the envelope at $r_d$. As shown by the right panel of Figure \ref{fig:shape}, however, all the solutions with different $\Lambda$ values are self-similar away from the base and actually have identical physical sizes in units of $r_{\rm s}$ (defined in Equation \ref{eqn:rs}). The height of the cavity is found to be $Z_{\rm max} \sim 2.5\,r_{\rm s}$, while the maximum cylindrical radius of the cavity is $R_{\rm max} \sim 0.16\,r_s$. It is important to note that the self-similarity of these solutions breaks down near the base, where the relevant scaling length remains $r_{\rm d}$ for all $\Lambda$.

\begin{figure*}[ht]
    \begin{center}
    \includegraphics[width=0.7 \columnwidth]{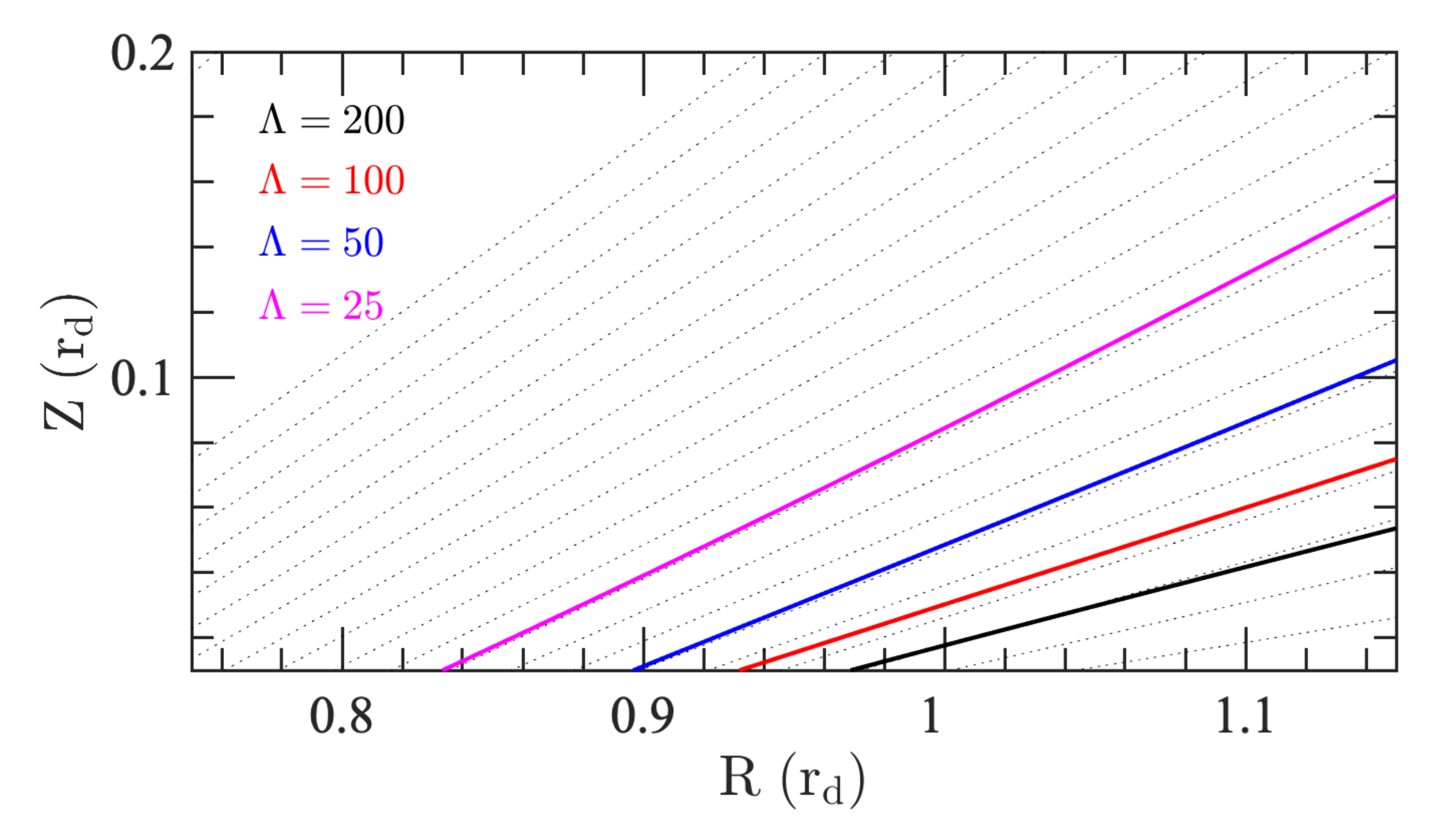}
    \vspace{-10 pt}
	\end{center}
    \caption{Cavity shapes near the base for an isotropic wind colliding with an \citet{ulrich76} infalling envelope for fixed $\lambda=1/2$ and varying $\Lambda$. The dashed lines represent streamlines for material flowing within the infalling envelope.
    }
    \label{fig:base}
\end{figure*}

\begin{figure*}[ht]
    \begin{center}
    \includegraphics[width=0.45 \columnwidth]{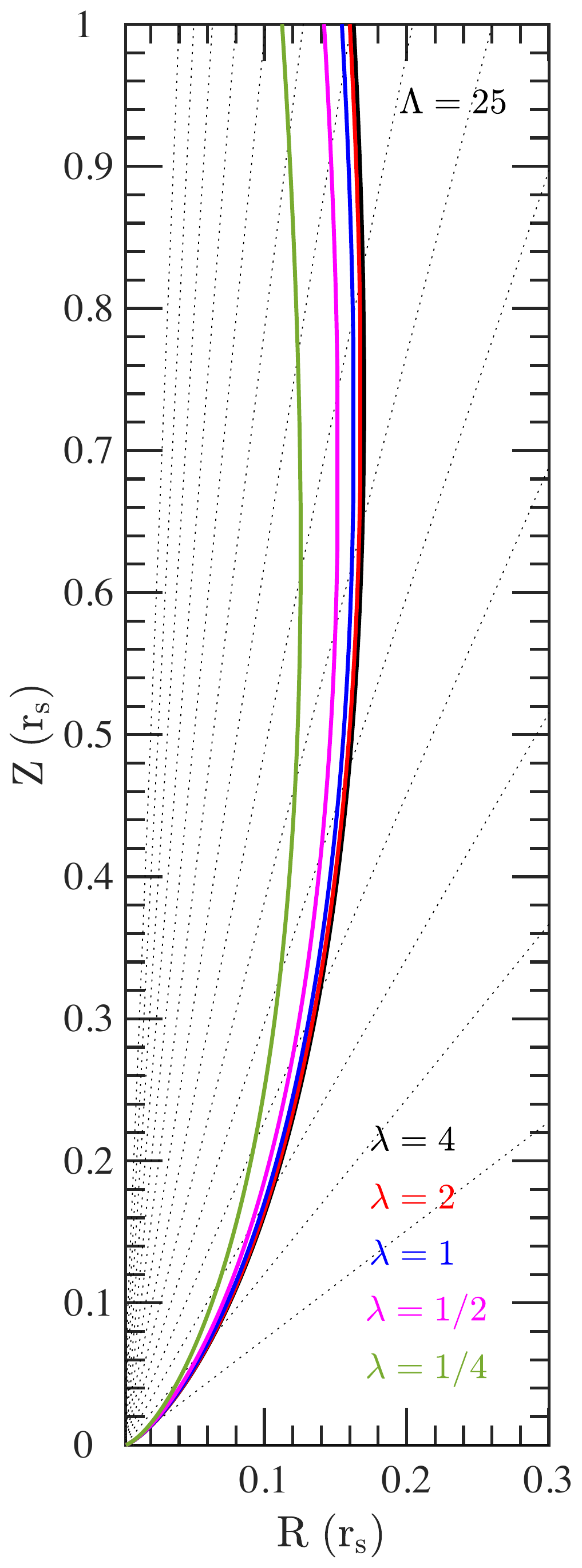}
    \vspace{-10 pt}
	\end{center}
    \caption{  
    Cavity profiles for fixed $\Lambda = 25$ and varying $\lambda$ showing breakout solutions when $\lambda  > 0.2$. The dashed lines represent streamlines for material flowing within the infalling envelope onto the disk of radius $r_d = 2\, r_s/\Lambda^2$. As discussed in the text, in units of $r_s$ the cavity shape becomes fixed for any $\Lambda$ at large $\lambda$. 
    }
    \label{fig:profile}
\end{figure*}

The location and incidence angle at which the cavity intersects the mid-plane for all these solutions is set by requiring that the cavity interface be tangent to the infalling envelope streamlines. Figure \ref{fig:base} shows in detail the cavity shapes near the disk surface, as well as the orientation of the infalling streamlines from the envelope for the cases investigated in Figure \ref{fig:shape}. From the figure, it is clear that the smaller $\Lambda$ solutions intersect the base somewhat interior to the larger $\Lambda$ solutions. These solutions therefore result in less interaction with the infalling envelope, as can be seen in the figure by noting the trajectories of the envelope streamlines.

\begin{figure}[ht]
    \begin{center}
	\includegraphics[width=0.9	\columnwidth]{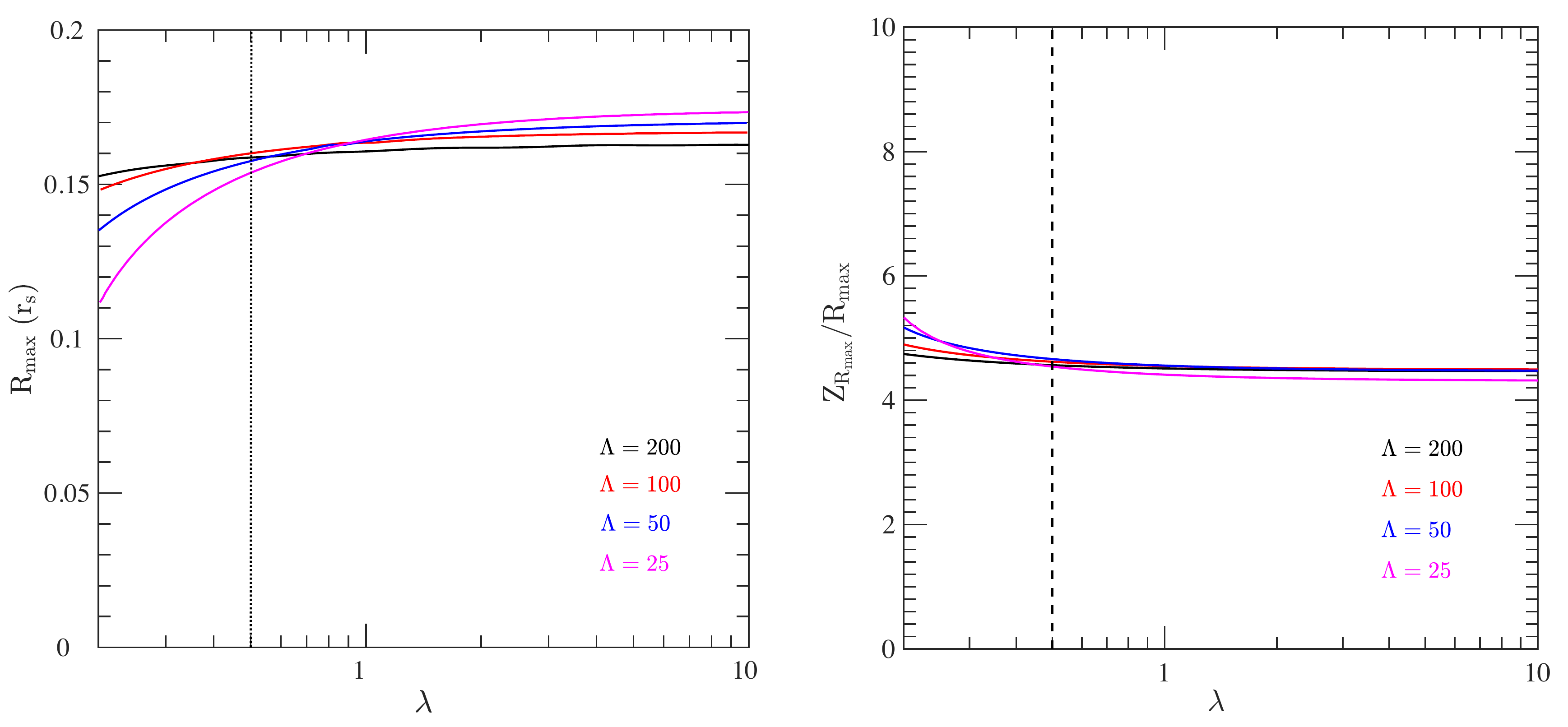}
	\end{center}
    \caption{Left: Maximum cylindrical radius, in units of $r_s$, for a cavity produced by an isotropic wind colliding with an \citet{ulrich76} infalling envelope for a range of $\lambda$ and $\Lambda$ values.
    We find steady breakout solutions even when $\lambda < 1/2$, to the left of the dotted vertical lines.
    Right: Ratio of the height of the cavity to the cylindrical radius of the cavity, measured at the location where the cavity is widest.}
    \label{fig:width}
\end{figure}

Until now we have fixed $\lambda = 1/2$; however, breakout solutions should exist for other values of $\lambda$. In Figure \ref{fig:profile} we show the cavity shape for $\Lambda = 25$ and a variety of $\lambda$. At the base, the $\Lambda$ solutions are {\it independent} of $\lambda$ while the maximum cylindrical radius and height increase with increasing $\lambda$, quickly asymptoting to a fixed solution. Furthermore, in Figure \ref{fig:width} we plot, as a function of $\lambda$, both the maximum cylindrical radius $R_{\rm max}$ of the cavity (in units of $r_s$) and the ratio of the height to the cylindrical radius at this widest point in the cavity, for each of the $\Lambda$ cases used in the previous figures. It is clear from these plots that for $\lambda > 1/2$ {\it all} the solutions become remarkably self-similar, with only a slight hint that the shapes are slightly broader for larger $\lambda$ and smaller $\Lambda$. We further note that steady breakout solutions are found even for values of $\lambda < 1/2$ (down to $\lambda \simeq 0.2$). These solutions differ from the  \citet{MCR04} trapped solutions which are required {\it a priori} to have $\partial r/\partial \theta = 0$ along the vertical axis,  creating a roundish ``cap" strongly confined by infall ram pressure when $\lambda < 1/2$. Instead, our $0.2 \lesssim \lambda < 1/2$ cavities maintain flame-like shapes, such that infall ram pressure at the tip is strongly reduced by the highly oblique incidence there, and allows for breakout.

The fact that our numerical solutions are not highly dependent on the initial location of the interface at the disk surface (see Appendix A), nor on the exact values of $\lambda$ or $\Lambda$, confirms that they are stable equilibrium solutions, as expected when the confinement is dominated by envelope pressure. 

\section{Flows Along the Cavity Wall}
\label{sec:flow}

The shape of the cavity wall as a function of the two defining input parameters $(\lambda, \Lambda)$ was shown in Section \ref{sec:shape} to be close to self-similar as long as $\lambda > 1/2$, especially at large distance from the intersection with the mid-plane. Thus it is reasonable to expect that the flow of deflected material from either the wind or envelope side of the cavity wall can also be described in terms of a single simplified parametrization, with small deviations as a function of  $\Lambda$ due primarily to the slightly varying physical situation near the disk surface.

\begin{figure}[ht]
    \begin{center}
    \includegraphics[width=0.8
    \columnwidth]{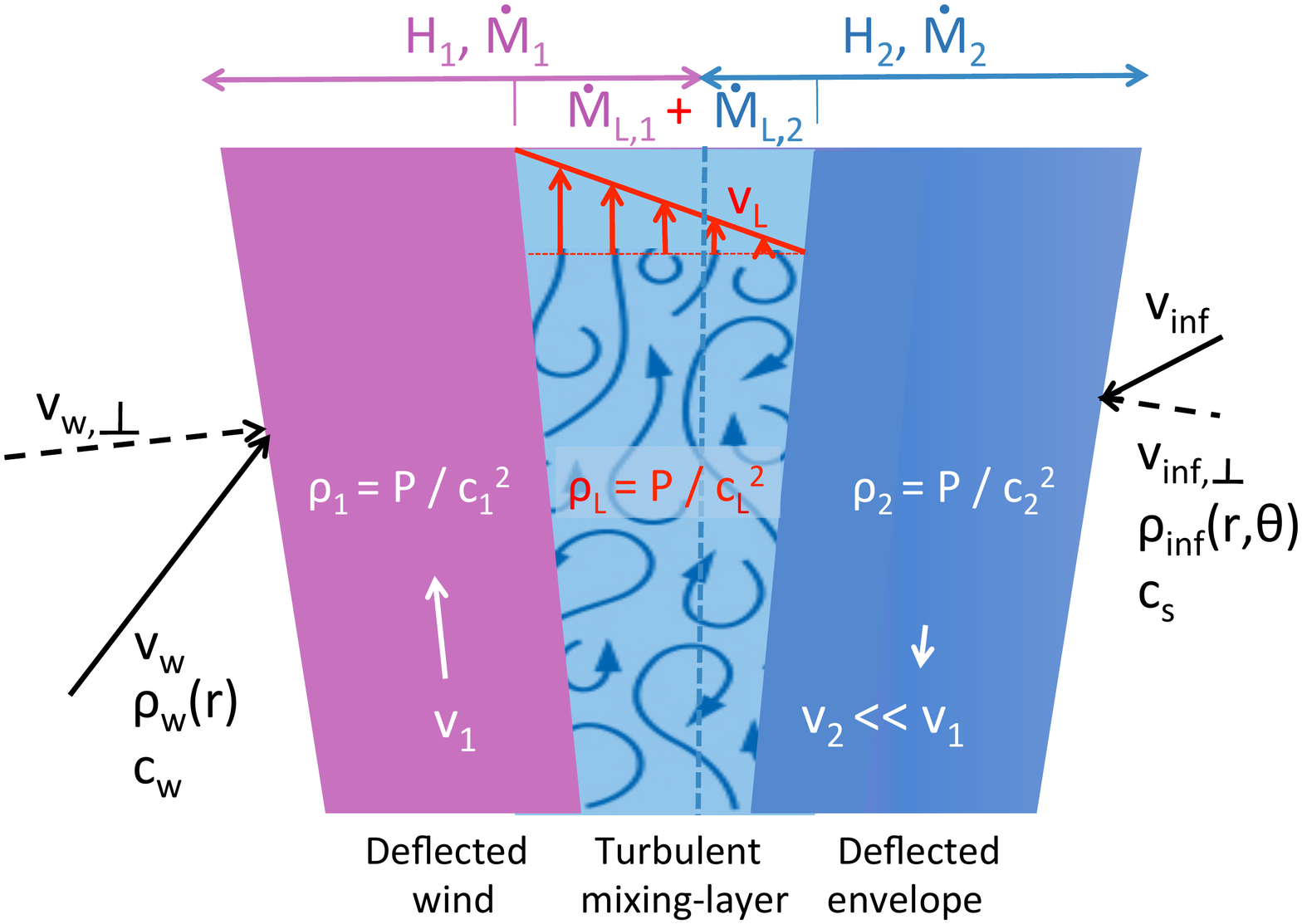}
	\end{center}
    \caption{Schematic showing the various layers along the cavity wall, in which deflected wind material moves upward and deflected envelope material moves downward. A central turbulent mixing-layer with a linear ``Couette-type" velocity gradient may grow between these two layers (\cf\ Section \ref{sec:flow:mix}). Mathematical notations used in the text for the velocity, mass-flux, sound speed, and density in each part of the flow are also shown for easy reference.}
    \label{fig:cartoon}
\end{figure}

In order to keep track of the various flows, we divide the cavity wall surface into three components. We present a schematic of the various regions in Figure \ref{fig:cartoon}.  First, there is the deflected shocked wind (denoted in the text by a subscript 1) that travels upward, parallel to and on the inside of the cavity surface. Second, there is the deflected infalling envelope (denoted in the text by a subscript 2) which travels downward, parallel to and on the outside of the cavity surface. In the absence of mixing across the surface, these two flows remain independent and can be fully described at each location by a mass and momentum flux explicitly determined by integration along the surface \citep[see equations {[2-5]} in][]{MJH09}. 
Third, a turbulent mixing-layer (denoted in the text by a subscript $L$) in which slow moving deflected envelope material is entrained upwards by the fast deflected wind, can develop at the contact discontinuity between these two flows.

\begin{figure}[ht]
    \begin{center}
    \includegraphics[width=0.9
    \columnwidth]{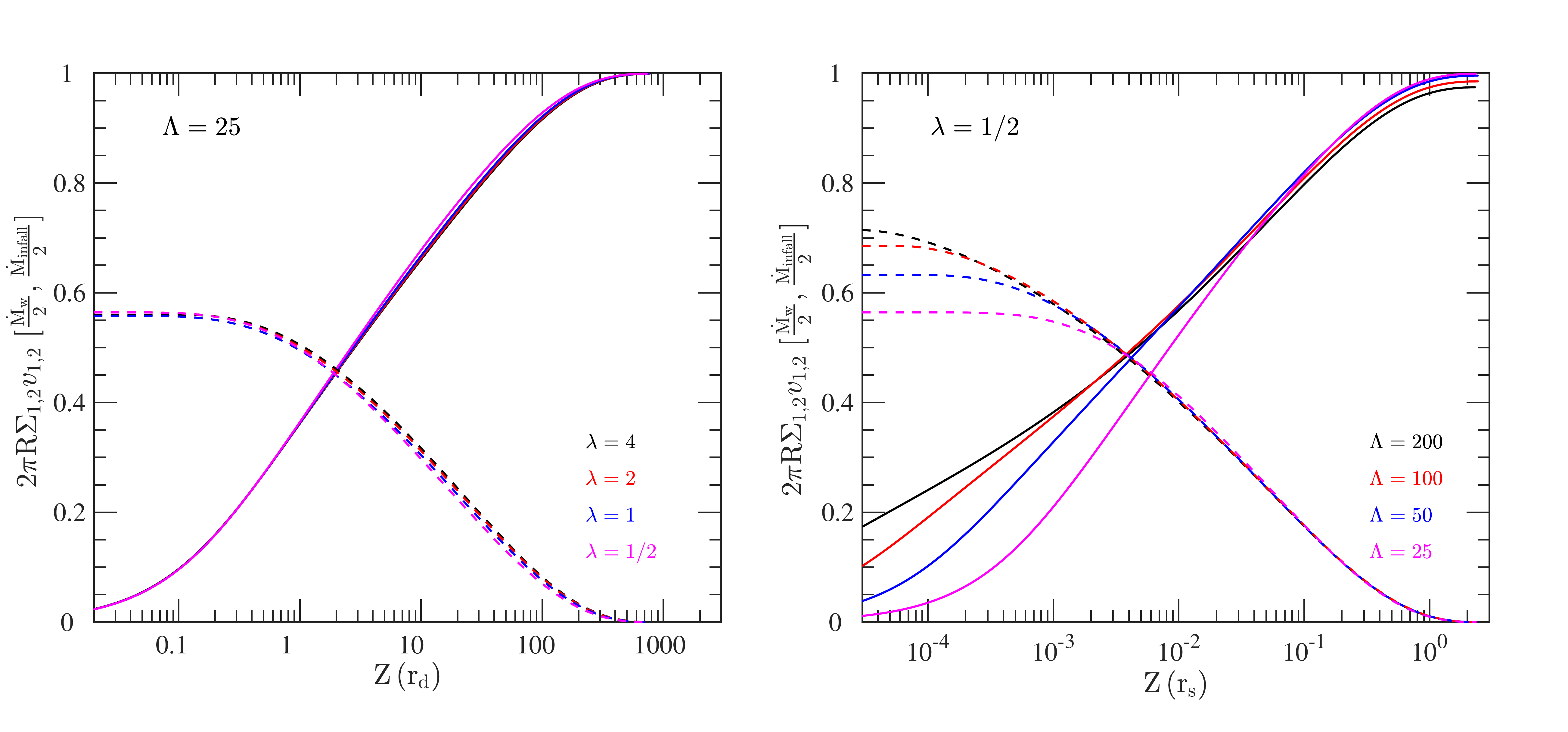}
	\end{center}
    \caption{Mass flow along both the interior (upward: solid line) and exterior (downward: dashed line) layers. The solutions shown are for one hemisphere only. The y-axis is scaled independently for the downward and upward flows. The left panel shows results with $Z$ in units of $r_{\rm d}$ for a variety of $\lambda$ while fixing $\Lambda = 25$. The right panel shows results with $Z$ in units of $r_{\rm s}$ for a variety of $\Lambda$ while fixing $\lambda=1/2$.}
    \label{fig:mass}
\end{figure}

\subsection{Solutions Without a Mixing-Layer}
\label{sec:flow:nomix}

We begin with solutions in which there is no mixing between the upward and downward deflected flows. Figure \ref{fig:mass} plots the mass flux as a function of (scaled) height for both the interior ($\dot M_1$; upward) and exterior ($\dot M_2$; downward) layers. 
The results are shown for a single hemisphere, and scaled accordingly, as there is no explicit requirement within the model for symmetry about the disk plane. The right panel fixes $\lambda = 1/2$ and shows solutions for four values of $\Lambda$ whereas the left panel fixes $\Lambda$ and shows solutions for four values of $\lambda$. In these plots it is important to recognize that the mass-flux axes are scaled separately, with the out-flowing material $\dot M_1$ scaled to the total mass flux from the wind over a hemisphere and the infalling material $\dot M_2$ scaled to the mass infall rate from the envelope over a hemisphere. Thus, while all the solutions are self-similar well above the mid-plane the relative importance of the upward versus downward mass flux depends on both $\lambda$ and $v_{\rm w}/v_{\rm d}$ (see Eqn.~\ref{eqn:lambda}). In all cases the upward flowing surface asymptotes to the entire mass flux in the wind, as required, whereas the the cavity intercepts only a  fraction of the infalling envelope, missing that part which lands on the disk surface between the cavity foot-point and $r_{\rm d}$. The trend with $\Lambda$, seen in the right panel, can therefore be understood as a direct consequence of the fact that smaller $\Lambda$ solutions intercept the disk closer to the central source (see Figure \ref{fig:base}). As shown in the left hand panel, {\it all} solutions at fixed $\Lambda$ are approximately self-similar modulo the wind and infall mass flux scaling. 

\begin{figure}[ht]
    \begin{center}
    \includegraphics[width=0.9 \columnwidth]{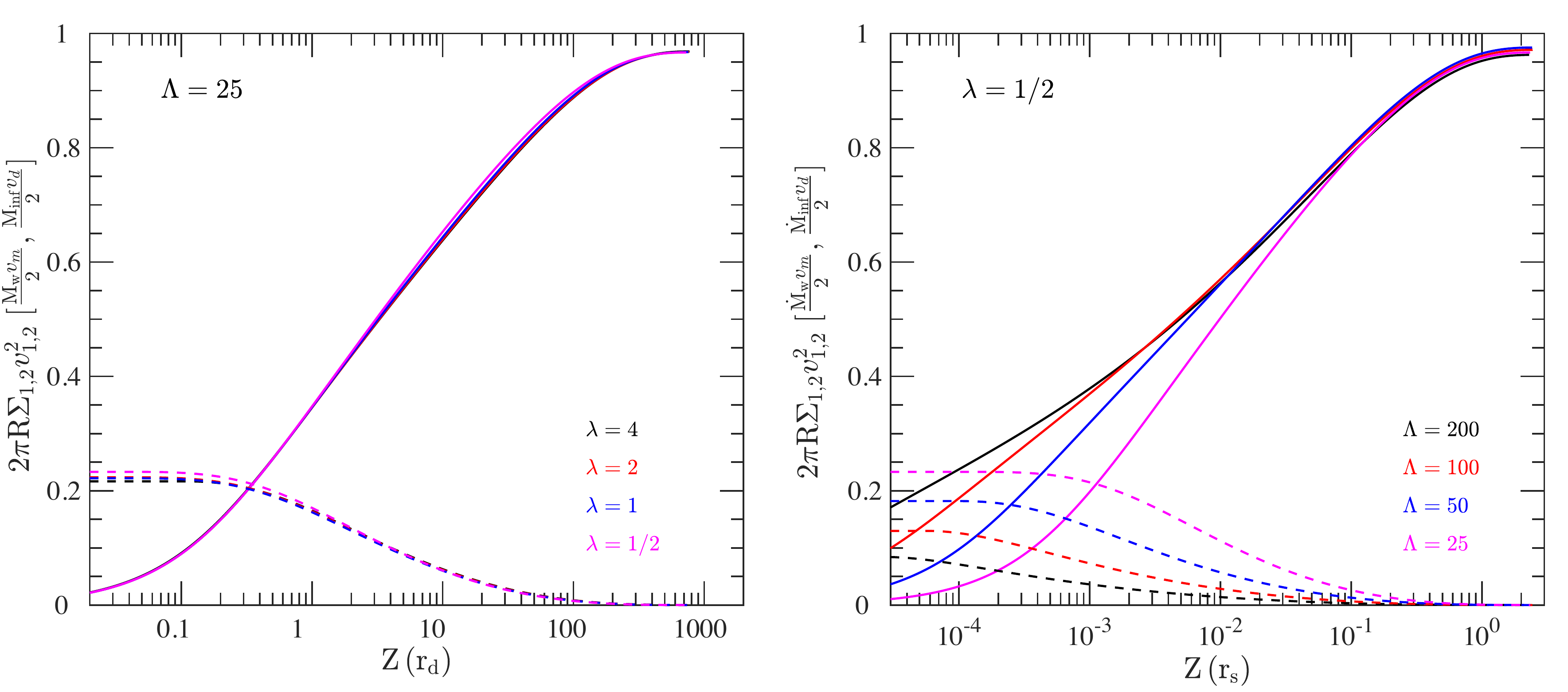}
	\end{center}
    \caption{Momentum flow along both the interior (upward: solid line) and exterior (downward: dashed line) layers. The solutions shown are for one hemisphere only. The y-axis is scaled independently for the downward and upward flows. The left panel shows results with $Z$ in units of $r_{\rm d}$ for a variety of $\lambda$ while fixing $\Lambda = 25$. The right panel shows results with $Z$ in units of $r_{\rm s}$ for a variety of $\Lambda$ while fixing $\lambda = 1/2$. 
    }
    \label{fig:mom}
\end{figure}

A similar set of solutions is found for the momentum flux as a function of height along the interior ($\dot \Pi_1 = \dot M_1\,v_1$; outward) and exterior ($\dot \Pi_2 = \dot M_2\,v_2$; inward) cavity layers, as shown in Figure \ref{fig:mom}. The quantities shown in the figure are scaled independently to the fiducial  momentum flux in the wind and infalling material. The relative scaling between these quantities, however, is explicitly $\lambda$ (see Eqn.~\ref{eqn:lambda}) and thus it is apparent that the downward momentum flux in the outer layer is always much less than the upward momentum flux in the inner layer except extremely close to the base. The left hand panel, again, shows that the solutions are approximately self-similar over a wide range of $\lambda$ for fixed $\Lambda$. 

\begin{figure}[ht]
    \begin{center}
    \includegraphics[width=0.98 \columnwidth]{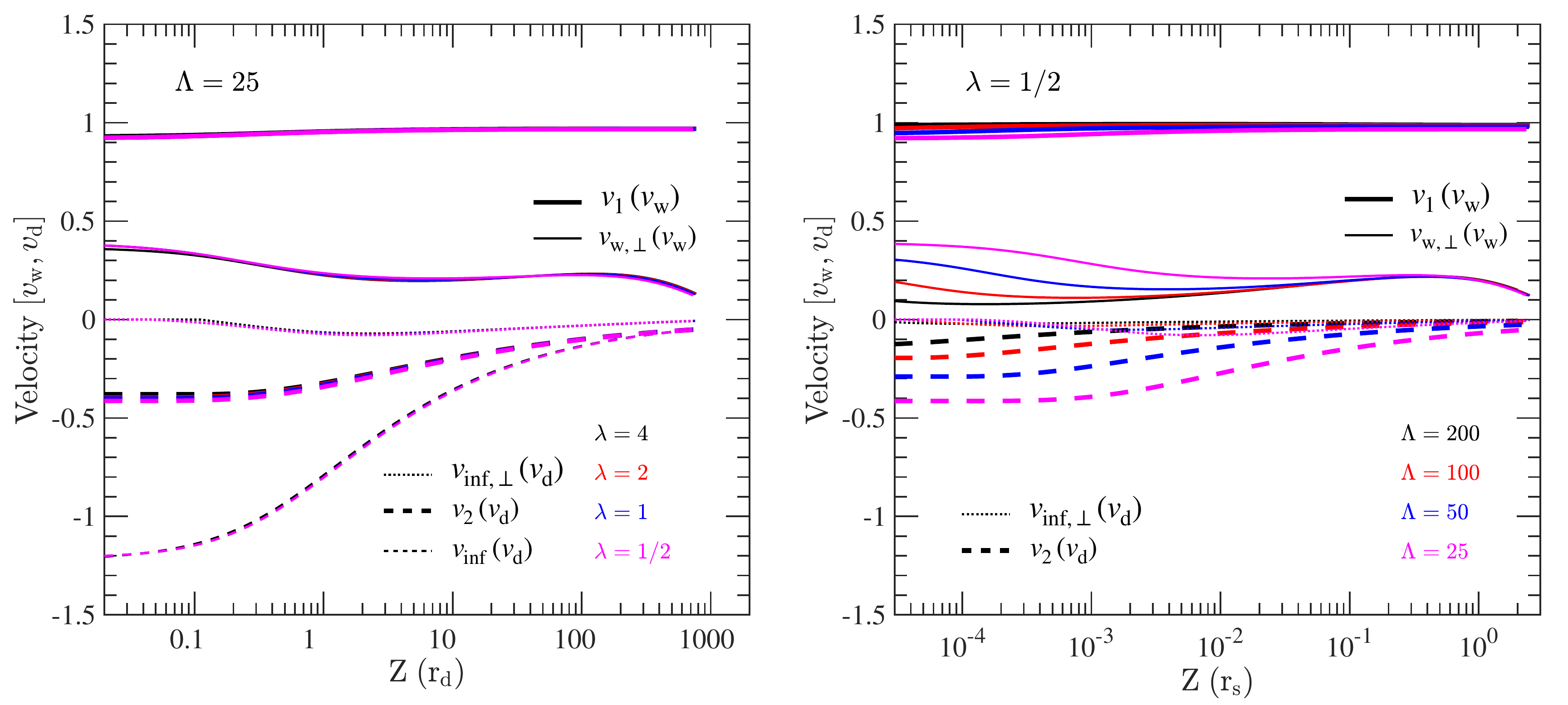}
	\end{center}
    \caption{Mean velocity along both the interior (upward: top two lines) and exterior (downward: bottom lines) layers. The solutions shown are for one hemisphere only. The y-axis is scaled independently for the upward and downward flows. The left panel shows results for a variety of $\lambda$ while fixing $\Lambda = 25$ with $Z$ in units of $r_{\rm d}$ while the right panel shows results for a variety of $\Lambda$ while fixing $\lambda = 1/2$ with $Z$ in units of $r_{\rm s}$.
    }
    \label{fig:vel}
\end{figure}

Finally, the mean velocities ($v_1, v_2$) for the two deflected flows along the surface can be calculated directly from the ratio of the respective momenta and mass fluxes. These are plotted in Figure \ref{fig:vel}. The quantities shown in the figure are scaled independently to $v_{\rm w}$ and $v_{\rm d}$.
The wind typically intersects the surface at an acute angle and thus the majority of the momentum from the wind is deposited in the deflected flow rather than contributing ram pressure to support the surface against the infalling envelope. Therefore, the magnitude of the velocity of the deflected wind along the surface remains near $v_{\rm w}$, asymptotically approaching $v_{\rm w}$ at large heights.  The infalling envelope, however, is almost stationary at large distances from the mid-plane and thus the deflected material at large heights shows little movement downward along the surface. The infalling velocity increases dramatically near the base, deeper in the potential well of the central object, and thus the downward velocity within the deflected envelope increases toward the mid-plane. Note that, the velocity $v_2$ along the external shell surface (deflected envelope) is always a tiny fraction of the upward velocity $v_1$ along the inner shell surface (deflected wind), as long as $v_{\rm w}/v_{\rm d} \gg 1$.

\begin{figure}[t]
    \begin{center}
	\includegraphics[width=1 \columnwidth]{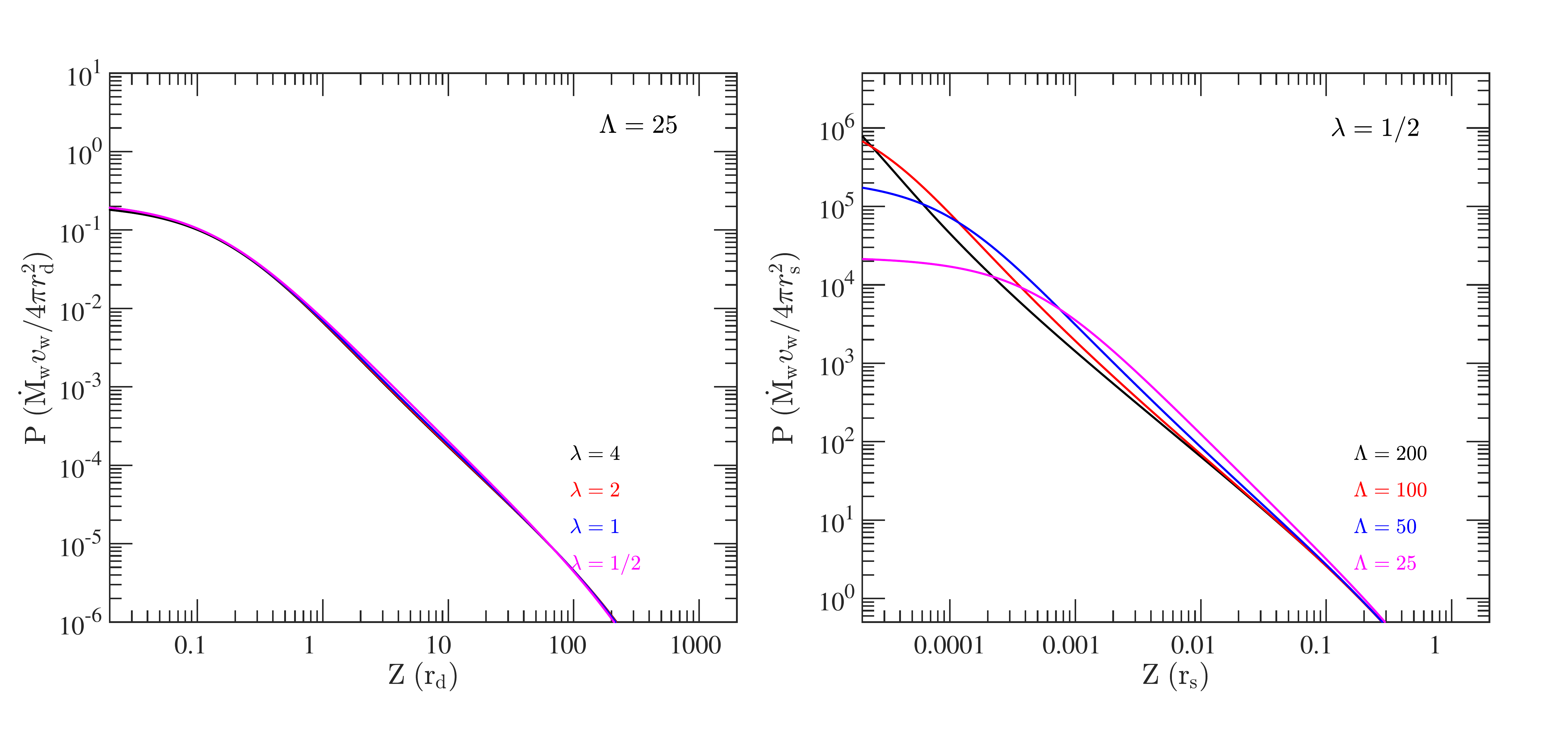}
	\end{center}
	\vspace{-15pt}
    \caption{Normalized pressure along the boundary surface of the cavity. The left panel shows results for a variety of $\lambda$ while fixing $\Lambda=25$ with $Z$ in units of $r_d$. The right panel shows results for a variety of $\Lambda$ while fixing $\lambda=1/2$ with $Z$ in units of $r_s$.
    }
    \label{fig:npml}
\end{figure}

One more calculation is required in order to complete the analysis of the flows in the absence of partial entrainment. For consistency with the assumptions, it is necessary to show that the shocked wind and shocked envelope layers remain thin in comparison to the radius of the cavity. To accomplish this we first determine the pressure confining the shocked layers (Figure \ref{fig:npml}). We note that on large scales the pressure may be well approximated by
\begin{equation}
P(Z) \simeq \left( \frac{\dot M_w v_{rm w}}{4 \pi r_s^2} \right)\, \left( \frac{0.2r_s}{Z} \right)^{1.5}
\simeq 0.1\, \left( \frac{\dot M_w v_{\rm w}}{4 \pi r_s^{0.5}} \right)\, Z^{-1.5}.
\label{eqn:pressure}
\end{equation}

Next, we compute the ratio of the thickness of the layers $H_1$ and $H_2$
against the radial extent of the cavity, as a function of location along the cavity. In detail, the thickness of each deflected layer is equal to the surface density divided by the mass density within the layer. In the thin shell approximation, the density within each layer is set by pressure equilibrium through $P = \rho_1 c_1^2 = \rho_2 c_2^2$, where $c_{1,2}$ are respectively the sound speed in the deflected wind and the deflected infalling envelope layer. Since $P$ is proportional to the ram pressure from the wind, the scalings for the relative thickness of the two deflected layers simplify to 
\begin{equation}
  \frac{H_1}{R}  \propto  \left( \frac{\dot{M}_{\rm w}}{v_{\rm w}} \right)
  \left( \frac{\dot{M}_{\rm w} v_{\rm w}}{c_1^2} \right)^{-1} \propto \left( \frac{v_{\rm w}}{c_1} \right)^{-2},
\label{eqn:H1}
\end{equation}
and
\begin{equation}
  \frac{H_2}{R} \propto  \left( \frac{\dot{M}_{\rm inf}}{v_{\rm d}} \right) 
  \left( \frac{\dot{M}_{\rm w} v_{\rm w}}{c_2^2} \right)^{-1} \propto \left( \frac{c_s}{c_2} \right)^{-2}\, \Lambda^{-1}.
\label{eqn:H2}
\end{equation}

\begin{figure}[th]
    \begin{center}
	\includegraphics[width=1 \columnwidth]{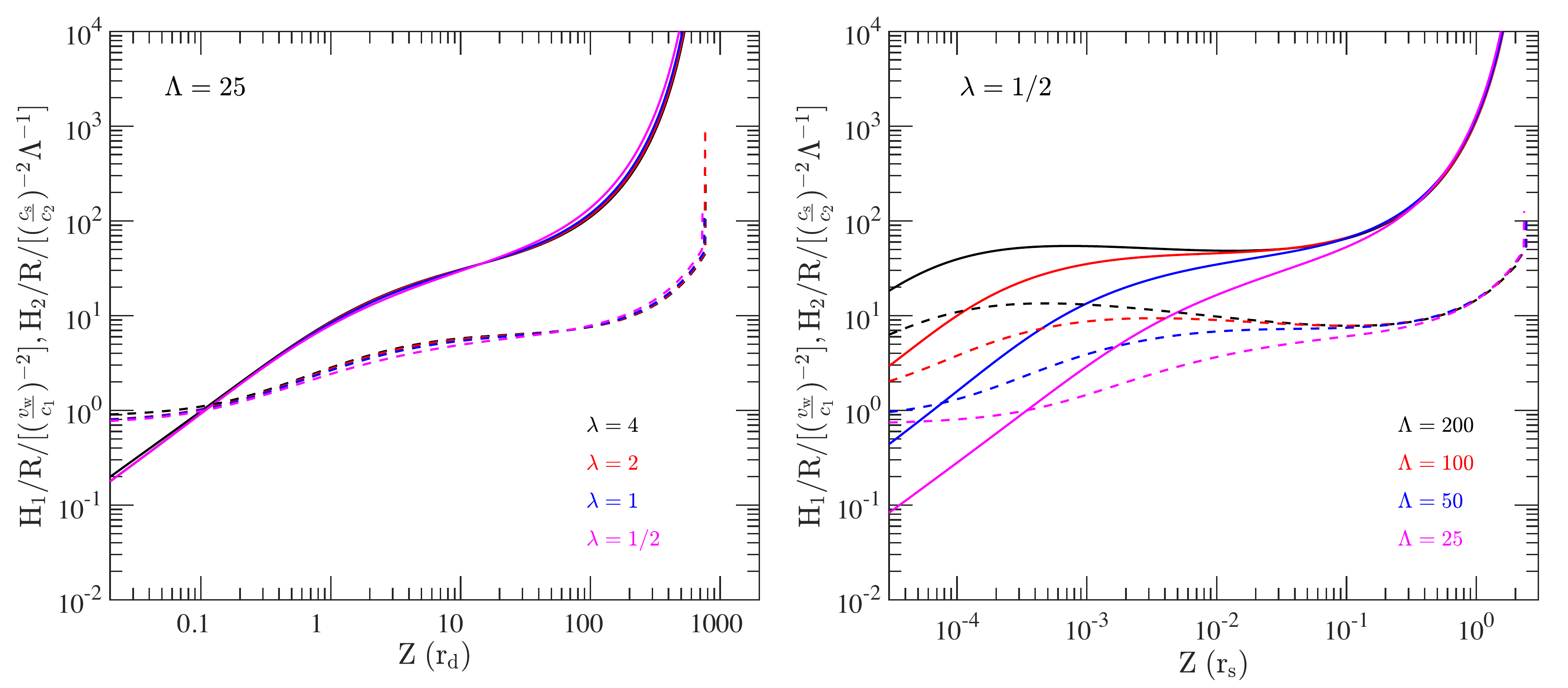}
	\end{center}
    \vspace{-10pt}
    \caption{Relative thickness of both the interior (upward: solid line) and exterior (downward: dashed line) layers as a function of height. The ratio $H/R$ is normalized in each case to the relevant scaling derived in Eqns.~\ref{eqn:H1}, \ref{eqn:H2}. The left panel shows results for a variety of $\lambda$ with $Z$ in units of $r_{\rm d}$ while fixing $\Lambda = 25$. The right panel shows results for a variety of $\Lambda$ with $Z$ in units of $r_{\rm s}$ while fixing $\lambda = 1/2$ . 
    }
    \label{fig:thick}
\end{figure}

Figure \ref{fig:thick} presents the relative thickness $H/R$ of the two layers, normalized by their respective relevant scaling in each case. Except near the top of the cavity, where the solution  converges back toward the axis of rotation, the deflected wind and envelope layers remain thin provided $v_{\rm w}/c_1 > 10$ and $\Lambda \, (c_s/c_2)^2> 10$, respectively. Furthermore, Figure \ref{fig:vel} shows that the deflected wind suffers oblique shocks with speed $v_s = v_{\rm  w,\perp} \simeq 0.2\,v_{\rm w}$. Using the general expression for the maximum temperature reached behind a hydrodynamical shock of speed $v_s$ to set an upper limit on $c_1$, it may then be determined that the condition for the shocked wind layer to remain thin is equivalent to $v_{\rm w}/c_{\rm w} > 10$, with $c_{\rm w}$ the isothermal sound speed in the wind.

Similarly, Figure \ref{fig:vel} also shows that the deflected envelope undergoes only small velocity jumps $v_s = v_{\rm  inf,\perp} < 0.1 v_{\rm  d}$. Since we will always have $c_2 \geq c_s$, a conservative condition ensuring that the deflected envelope layer will remain thin, regardless of the value of $v_d$, is simply that $\Lambda > 10$. This conservative condition on $\Lambda$ becomes unnecessary, however, if $v_d$ is large enough for the velocity jumps to remain supersonic everywhere along the cavity wall. Using the expression for the maximum temperature behind a shock at $v_s$ to set an upper limit on $c_2$, and rewriting the condition for a thin envelope layer as $\lambda > 5\, (c_2/v_d)^2 $\, (using  Eqn.~\ref{eqn:Lambda}), we find that the inequality is then automatically fulfilled under our wind breakout condition $\lambda > 0.2$, regardless of the value of $\Lambda$.


The combined results presented in Figures \ref{fig:mass} -- \ref{fig:thick} reveal that, in the absence of mixing, the deflected wind mass flowing upward along the boundary surface will be similar to the total mass flowing in the wind, $\dot M_{\rm w}$, and that the magnitude of the momentum in this deflected flow will also be close to the wind momentum flux $\dot M_{\rm w} v_{\rm w}$.

Due to the large aspect ratio of the cavity, the bulk of the deflected wind flows roughly perpendicular to the disk and at a high velocity, $v_1 \sim v_{\rm w}$.  Alternatively, if mixing takes place between the momentum-rich outward flowing layer and the mass-rich infalling layer, the internal velocity structure of this turbulent mixing-layer should be significantly differentiated, as for example through a linear velocity gradient such as occurs in a Couette flow \citep[e.g.][]{rcc95}. Such an occurrence will naturally produce a wider spread of velocities between $v_2$ and $v_1$. 
 
\subsection{Solutions with a Mixing-Layer}
\label{sec:flow:mix}

A general formalism for the growth of a turbulent mixing-layer between two  axisymmetric flows was derived by \citet{rcc95}. Their main formulae included some ambiguities and typographical errors, and are therefore reproduced in corrected form in Appendix \ref{ap:mixequ}. These authors assume that within the mixing-layer there is both a fixed temperature, referred to by its sound speed $c_L$, and a fixed pressure $P$ across the layer, varying only as a function of position along the flow. Across the layer they further assume that the velocity varies linearly, as in a Couette flow, bounded by the velocities of the fast $v_1$ layer and slow $v_2$ layer (see Figure \ref{fig:cartoon}). The coupled equations for the change in mass flux and momentum flux within the mixing-layer due to entrainment across the inner and outer boundaries are then solved so as to determine the entrainment required across each bounding surface in order to maintain the imposed Couette conditions. 

As detailed by \citet{rcc95}, entrainment occurs in two ways (see their equations 1 and 2): through the geometrical growth of the mixing-layer, intercepting a fraction of the flows on either side, and through ``turbulent entrainment" on the slow-moving side (here the deflected envelope) as it is dragged into the mixing-layer by the fast-moving side. In this paper, we follow the \citet{rcc95} prescription for the turbulent entrainment velocity $v_{\rm ent} = \alpha c_2^2/c_L$ (see their equation 3), with a constant turbulent mixing parameter $\alpha$.  For simplicity, we will further assume constant values of $c_L$, $c_1$ and $c_2$ at all positions.

At any location $x$ along the surface, the linear velocity gradient across the turbulent layer assures that the mass-weighted mean velocity within the mixing-layer is $\overline{v_L}(x) = \left[v_1(x) + v_2(x)\right]/2$. At the same time, the ratio of the momentum flux $\dot \Pi_L$ versus the mass flux $\dot M_L$ in the mixing-layer is skewed toward the higher velocities within the layer, such that
\begin{equation}
\frac{\overline{v_L^2}}{\overline{v_L}} \equiv \frac{\dot \Pi_L(x)}{\dot M_L(x)} = \frac{2}{3}\,\left(\frac
{v_1^3(x) - v_2^3(x)}
{v_1^2(x) - v_2^2(x)} \right).
\label{eqn:vbar2}
\end{equation}

To maintain this ratio as slow envelope material $\dot M_{L2}$ is turbulently entrained across the outer boundary surface, a significant amount of fast-flowing shocked wind material $\dot M_{L1}$ must also be entrained across the inner boundary, with the exact proportion set by the (changing) physical conditions along the surface. For the trivial case of a constant velocity, $v_1$, in the fast moving layer and no motion on the slow-moving side ($v_2 = 0$), Eqn.~\ref{eqn:vbar2} shows that the Couette flow requires ${\dot \Pi_L(x)}= (2v_1/3){\dot M_L(x)}$. Since zero momentum can be provided from the stationary side, this condition is met by
the mass entrainment rate from the fast moving wind side being exactly twice the mass entrainment rate from the stationary side \citep{cr91}. 

\begin{figure}[th]
    \begin{center}
	\includegraphics[width=1 \columnwidth]{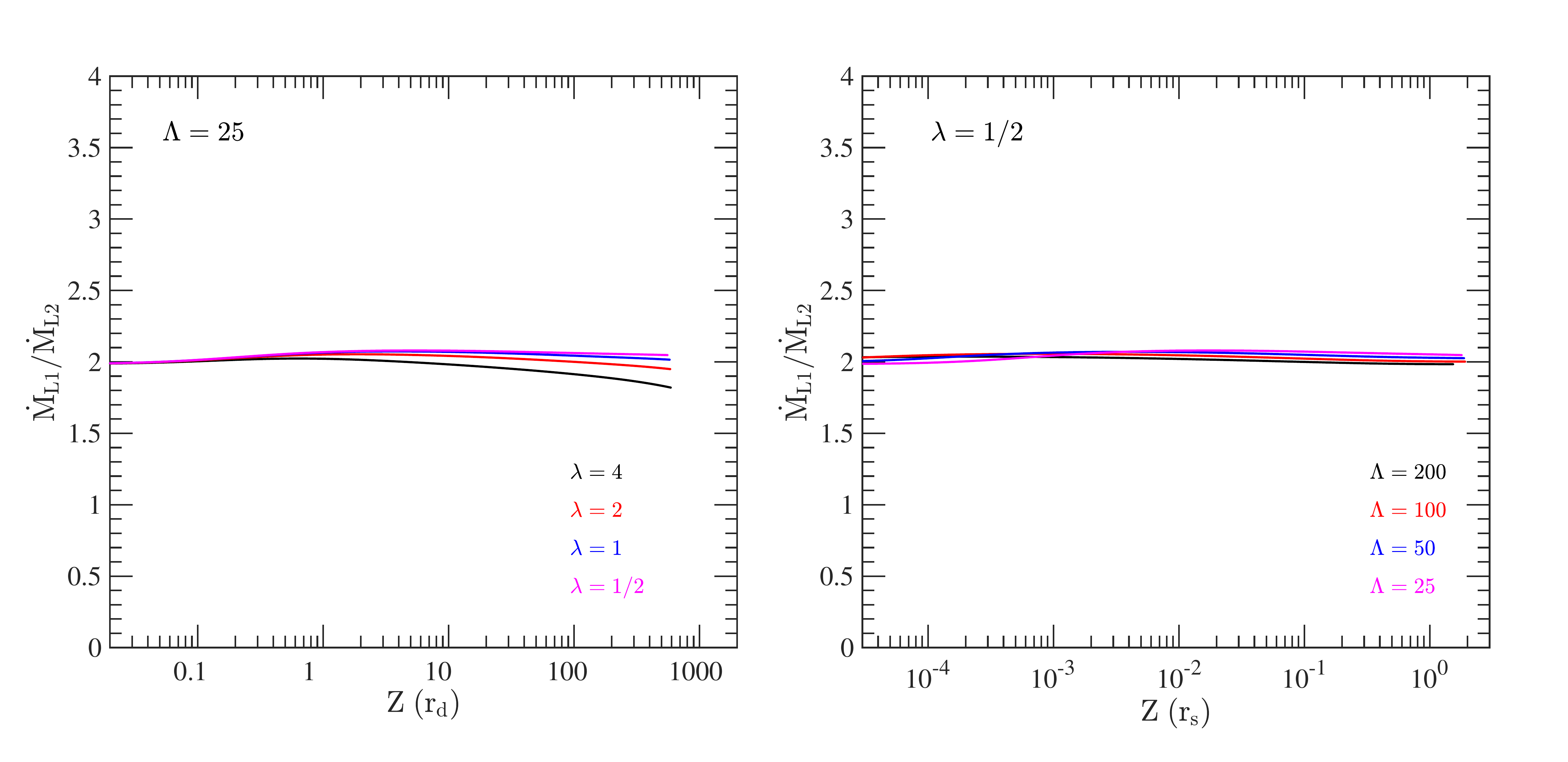}
	\end{center}
	\vspace{-10pt}
    \caption{Computed ratio of the mass flux entrained within the mixing-layer from the fast moving shocked wind material, $\dot M_{L1}$, versus the turbulent entrainment from the shocked envelope, $\dot M_{L2}$. Note that for all solutions the ratio remains close to 2, as expected to maintain a linear Couette velocity profile across the mixing-layer.
    }
    \label{fig:ratio}
\end{figure}

A careful consideration of the interface solutions presented in Section \ref{sec:shape} shows that for the self-similar cavities in this paper $v_1 \sim v_{\rm w}$ and $|v_2| \ll v_{\rm w}$, in all examined cases. This simplifies the general formulae described by \citet{rcc95}; however, the changing radius of curvature and the steadily dropping pressure across the calculated wind-envelope surface conspire such that the detailed solution for entrainment must be calculated numerically and separately for all parameter pairs ($\lambda$, $\Lambda$). Fortunately, despite this somewhat more complicated geometry, Figure \ref{fig:ratio} shows that for all pairs ($\lambda$, $\Lambda$) 
the mass-flux entering the mixing-layer from the fast-flowing wind side, $\dot M_{L1}$, remains close to twice the turbulent entrainment from the envelope side, $\dot M_{L2}$.

The self-similarity of the boundary location also provides, for each ($\lambda$, $\Lambda$) pair, a scaling relation for the efficiency of the turbulent mixing solutions in terms of the physical parameters $\dot M_{\rm w}, v_{\rm w}, c_L$, as well as $\alpha$. The turbulent entrainment into the mixing-layer from the slow-moving, envelope, side of the boundary, $\dot M_{L2}$, can be found by integrating the turbulent entrainment along the boundary surface, $x$.  That is:
\begin{equation}
    \dot M_{L2} = 2 \pi \int R(x) \rho_2(x)\, v_{\rm ent}\, dx,
\end{equation}
where $\rho_2(x)$ is the density of the deflected shocked envelope  along the outer boundary and $v_{\rm ent} = \alpha (c_2^2/c_L)$ is the parametrized entrainment velocity. This equation is exact for situations where the shocked ambient medium is static and remains an excellent estimate when $|v_2| \ll v_{\rm w}$. The result may be rewritten in terms of the pressure across the boundary surface
\begin{equation}
    \dot M_{L2} = 2 \pi \left(\frac{\alpha}{c_L}\right) \int R(x) P(x) dx,
\end{equation}
which further reduces by recognizing that the pressure at location $x$ along the surface is set explicitly by the ram pressure: $P(x) = a(x)\,\dot M_{\rm w}\,v_{\rm w}$ where $a(x)\propto \sin^2{\gamma}/r^2$ takes into account the varying angle of incidence between the isotropic wind and the boundary surface. Thus,
\begin{equation}
    \dot M_{L2} = 2 \pi \left(\frac{\alpha\,\dot M_{\rm w}\,v_{\rm w}}{c_L}\right) \int R(x) a(x) dx.
\end{equation}
Furthermore, given the almost fixed ratio between mass entrained into the mixing-layer from the deflected wind versus the deflected envelope (see Figure \ref{fig:ratio}), the scaling for $\dot M_{L1}$, and $\dot M_{L} = \dot M_{L1} + \dot M_{L2}$, should be the same as that for $\dot M_{L2}$. 

\begin{figure}[ht]
    \begin{center}
	\includegraphics[width=1 \columnwidth]{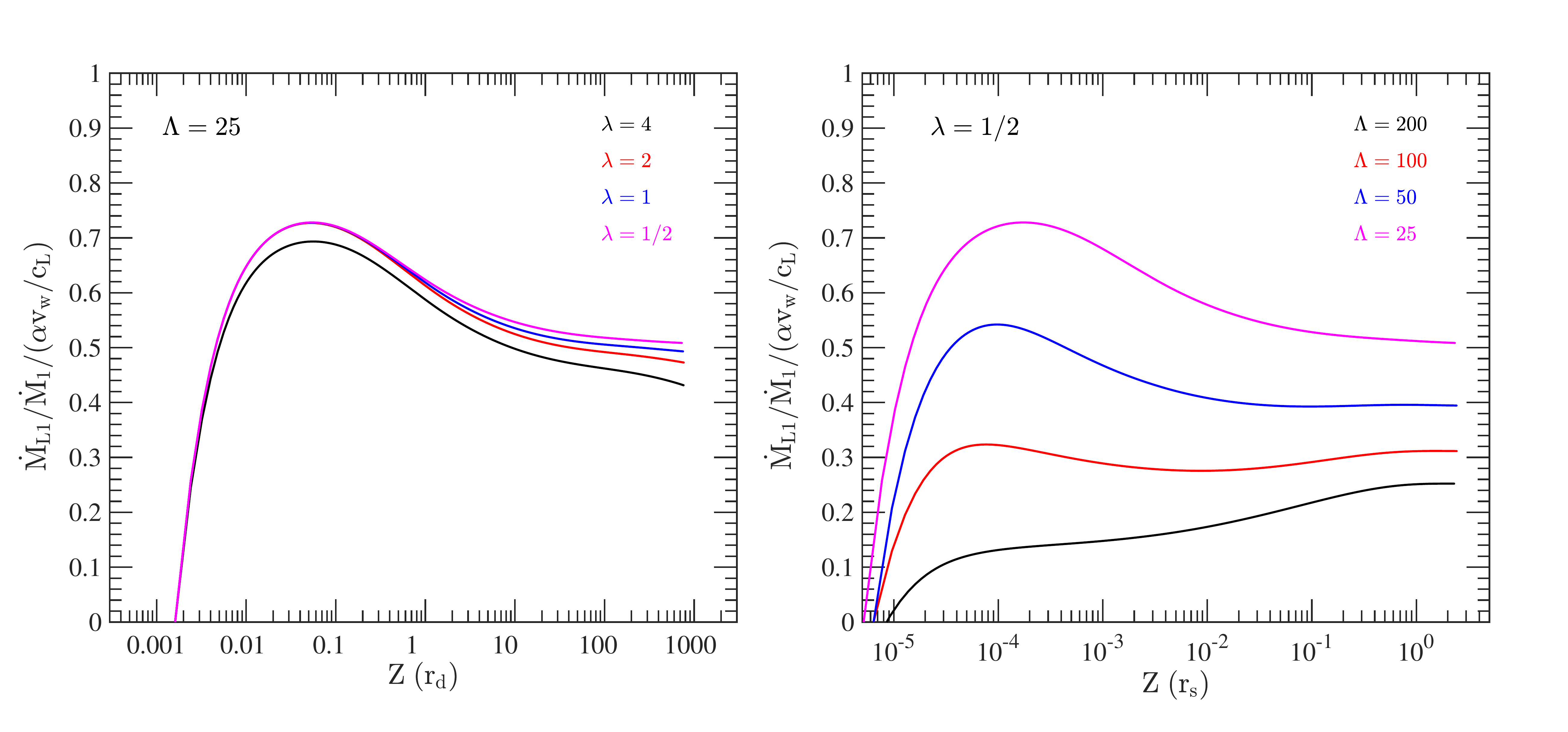}
	\end{center}
	\vspace{-15pt}
    \caption{Computed ratio of the cumulative material gained by the mixing-layer from the shocked wind ($\dot M_{L1})$ versus the available reservoir of shocked wind material including that already in the mixing-layer.
    }
    \label{fig:ml1}
\end{figure}

\begin{figure}[ht]
    \begin{center}
	\includegraphics[width=1 \columnwidth]{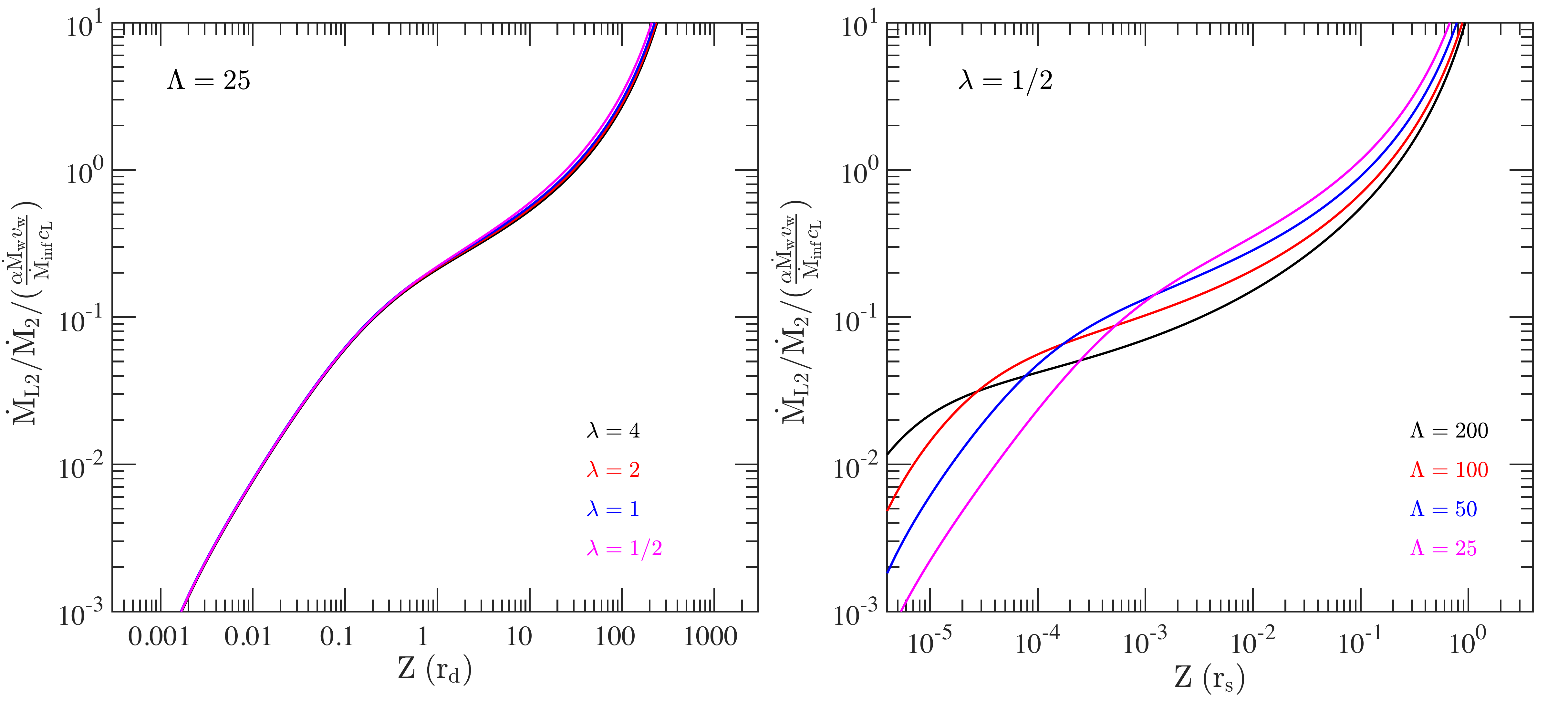}
	\end{center}
	\vspace{-15pt}
    \caption{Computed ratio of the cumulative material gained by the mixing-layer from the deflected envelope ($\dot M_{L2})$ versus the available reservoir of deflected envelope material including that already in the mixing-layer.
    }
    \label{fig:ml2}
\end{figure}

Utilizing this scaling as normalization, Figures \ref{fig:ml1} and \ref{fig:ml2} show the fraction of deflected wind and deflected envelope that is entrained into the mixing-layer as a function of height above the mid-plane, for a variety of ($\lambda$, $\Lambda$) pairs. Self-consistent solutions require that these fractions remains less than unity, otherwise the reservoir of shocked material flowing along the cavity walls is not large enough to feed material into the mixing-layer at our assumed rates. From Figure \ref{fig:ml1}, it is clear that for the wind side this constraint requires 
\begin{equation}
   \alpha \lesssim \alpha_{\rm max} \equiv \left(\frac{c_L}{v_{\rm w}}\right).
    \label{eqn:alpha}
\end{equation}
Similarly, for the envelope side (Figure \ref{fig:ml2}) the constraint is trivially met for the same physical parameters assuming $\dot M_{\rm w} < 0.5 \dot M_{\rm inf}$, except at extreme heights, $Z > r_{\rm s}$,  where the cavity shape converges to the axis of rotation.

\begin{figure}[t]
    \begin{center}
	\includegraphics[width=1 \columnwidth]{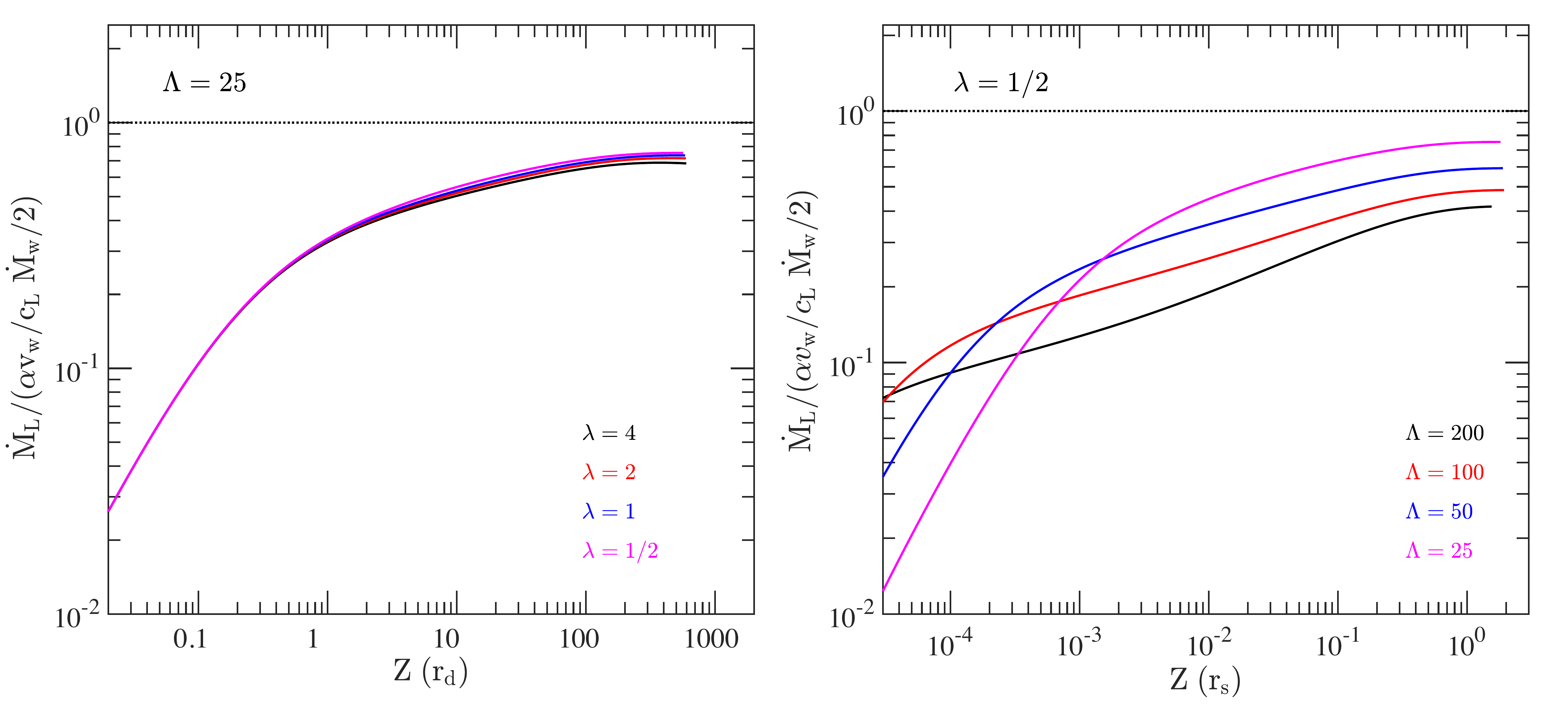}
	\end{center}
	\vspace{-15pt}
    \caption{Mass flux of the material within the mixing-layer.
    }
    \label{fig:mlt}
\end{figure}

Furthermore, combining the information in Figures \ref{fig:ml1} and \ref{fig:ml2}, and using the results of Figure \ref{fig:mass}, Figure  \ref{fig:mlt} shows that the total mass flux in the mixing-layer for one outflow cavity lobe is almost independent of the ($\lambda$, $\Lambda$) pair. Within a factor of a few,
the asymptotic value at high altitudes is
\begin{equation}
    \dot M_{L} \sim \left(\frac{\alpha\,v_{\rm w}}{c_L}\right) \frac{\dot M_{\rm w}}{2} = \left(\frac{\alpha}{\alpha_{\rm max}}\right) \frac{\dot M_{\rm w}}{2},
    \label{eqn:mlml}
\end{equation}
where $\alpha_{\rm max}$ is defined in Eqn.~\ref{eqn:alpha}. We note that if $\alpha > \alpha_{\rm max}$, our physical model will not entirely break down. The mixing-layer will simply grow until it eventually engulfs all of the deflected wind layer and $\dot M_{L}$ saturates at its maximum possible value of ${\dot M_{\rm w}}/{2}$. Without a fast laminar wind layer to enforce a Couette flow, however, the velocity field in the mixing-layer would become a gaussian velocity distribution peaked around a mean value $\simeq v_{\rm w}/2$, leading to line profiles much narrower than in a linear gradient Couette flow. 

\section{Calculation of Angular Momentum and Line Profiles}
\label{sec:profile}

\subsection{Angular Momentum of the Shocked Envelope and Mixing-Layer}
\label{sec:angular}

To compute the angular velocity of the shocked envelope, $v_{2, \phi}$, as well as the mixing-layer, $v_{L, \phi}$, we adopt the following equations (\cf\ equation 5 in \citealt{MJH09}),

\begin{equation}
    \rho_{\rm inf} (v_{\rm inf} \sin\,\theta_{\rm sa}) \Omega_{\rm inf, \phi} = \frac{\cos\beta }{R^3}\frac{\partial}{\partial R} (R^3\Sigma_2v_2\Omega_{2, \phi}),
    \label{eq.24}
\end{equation}
\noindent and
\begin{equation}
    \rho_2 (\alpha c_2) \Omega_{2, \phi} = \frac{\cos\beta }{R^3}\frac{\partial}{\partial R} (R^3\Sigma_{L}v_{L}\Omega_{L, \phi}),
    \label{eq.25}
\end{equation}

\noindent where $\Omega_{2, \phi}=v_{2, \phi}/(2\pi R)$ and $\Omega_{L, \phi}=v_{L, \phi}/(2\pi R)$. 

\begin{figure}[ht]
    \begin{center}
	\includegraphics[width=0.76 \columnwidth]{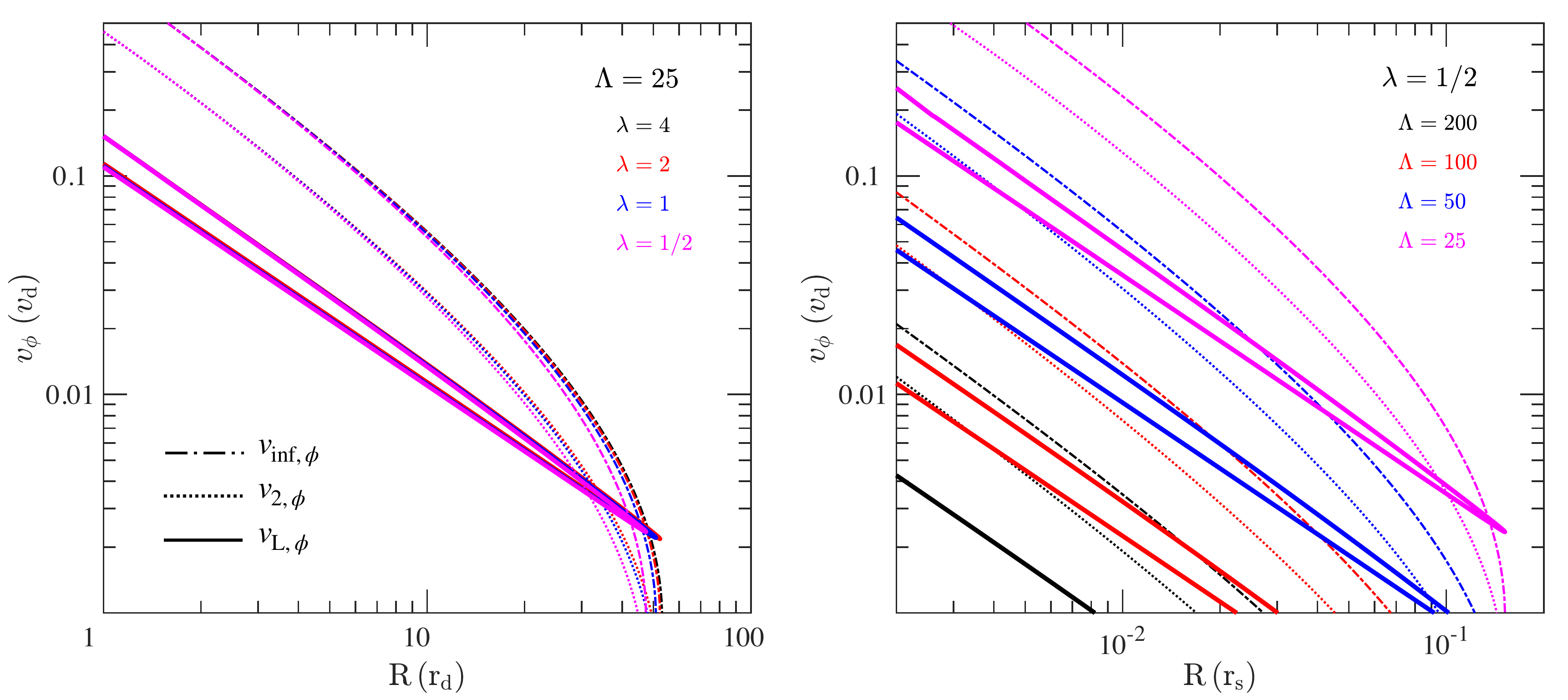}
	\includegraphics[width=0.76 \columnwidth]{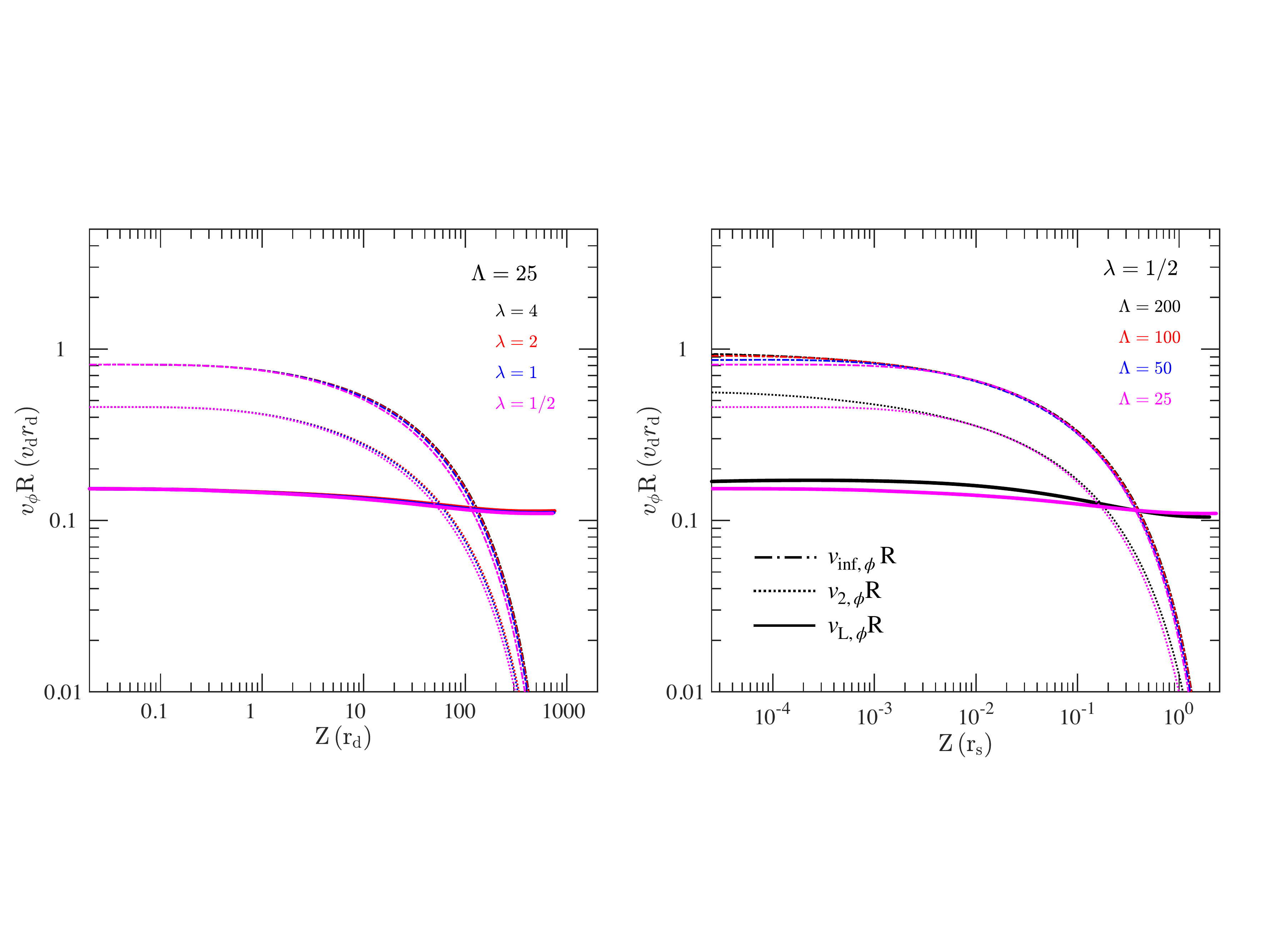}
	\end{center}
    \caption{The $\phi$ component of the infalling envelope at the interface (dot-dashed), in the shocked envelope gas (dotted) and in the gas within the central mixing-layer (solid). Top panels plot velocity while bottom panels plot specific angular momentum. Left panels compare results for fixed $\Lambda=25$ and varying $\lambda$ with $R$ in units of $r_d$. Right panels compare results for fixed $\lambda=1/2$ and varying $\Lambda$ with $R$ in units of $r_s$.
    }
    \label{fig:13}
\end{figure}

By integrating Eqn.~\ref{eq.24} over $R$ along the interface from the top of the cavity to the disk plane, we obtain $\Omega_{2, \phi}$, and hence $v_{2, \phi}$. Subsequently by  inputting the derived $\Omega_{2, \phi}$ into Eqn.~\ref{eq.25} and integrating it over $R$ from the disk plane upward along the interface,  we obtain $\Omega_{L, \phi}$, and hence $v_{L, \phi}$. In the top panels of Figure \ref{fig:13}, we show $v_{L, \phi}$ (solid lines), $v_{2, \phi}$, and $v_{\rm inf, \phi}$ as a function of $R$ for fixed $\Lambda$ and varying $\lambda$ (left panels), as well as for fixed $\lambda$ and varying $\Lambda$ (right panels). We can see that for $\Lambda > 20$, $v_{L, \phi}$ is no larger than a few tenths of $v_{\rm d}$ and therefore can be considered as negligible compared to the bulk velocity of the gas in the mixing-layer ($v_{L}\approx 0.5\, v_{\rm w}$) when determining the shell shape and computing the observed line profiles. 

The bottom panels of Figure \ref{fig:13} plot the specific angular momentum $R v_{\phi}$ for the same three velocity components as in the upper panels, as a function of $Z$. We can see that the specific angular momentum in the mixing-layer is virtually independent of $\lambda$ and $\Lambda$. Since twice as much material is entrained from the (non-rotating) wind side than from the envelope side, the initial value of the specific angular momentum at the base is one third of that in the deflected envelope. As gas is advected upwards in the mixing-layer, this rotating material gets mixed with deflected ambient material of smaller specific angular momentum. However, since most of the mass entrainment occurs at $Z \le 0.1\,r_s$, the specific angular momentum in the mixing-layer remains close to its initial value, $\simeq$ 0.15 $\, v_d \,r_d$.

\subsection{Mixing-Layer Line Profile}
\label{sec:line_profile}

Having developed a model for how material is entrained, the next step is to calculate the resulting line profile for direct comparison to observations. The model calculations in the preceding sections were all dimensionless and are thus applicable to essentially any type of entrainment irrespective of physical scale. Thus, with the appropriate scaling this model could be compared with outflows from protostars \citep[e.g.,][]{mottram14,Kristensen17} to extra-galactic outflows \citep[e.g.][]{aalto16, aalto17}.

The line profiles for the shocked wind layer and for the turbulent mixing-layer are generated independently  
under the optically thin assumption by computing the flux 
$dF_{\rm e}(v_{\rm LOS})_=\epsilon_{\rm e} dV$
emitted by each elementary volume element $dV$, where $\epsilon_{\rm e}$ is the emissivity per unit volume of the molecular line of interest. At every height, $z$, along the surface, the flux from each azimuthal interval, $d\phi$, is added to the velocity bin corresponding to the line-of-sight velocity
$v_{\rm LOS}$ of the volume element. 
For the shocked wind layer, the velocity  has a unique modulus $v_1$, dependent only on $z$, whereas for the mixing-layer, the flux is evenly distributed in velocity between 0 and $v_1$ prior to projection.

\begin{figure}[ht]
    \begin{center}
	\includegraphics[width=0.75 \columnwidth]{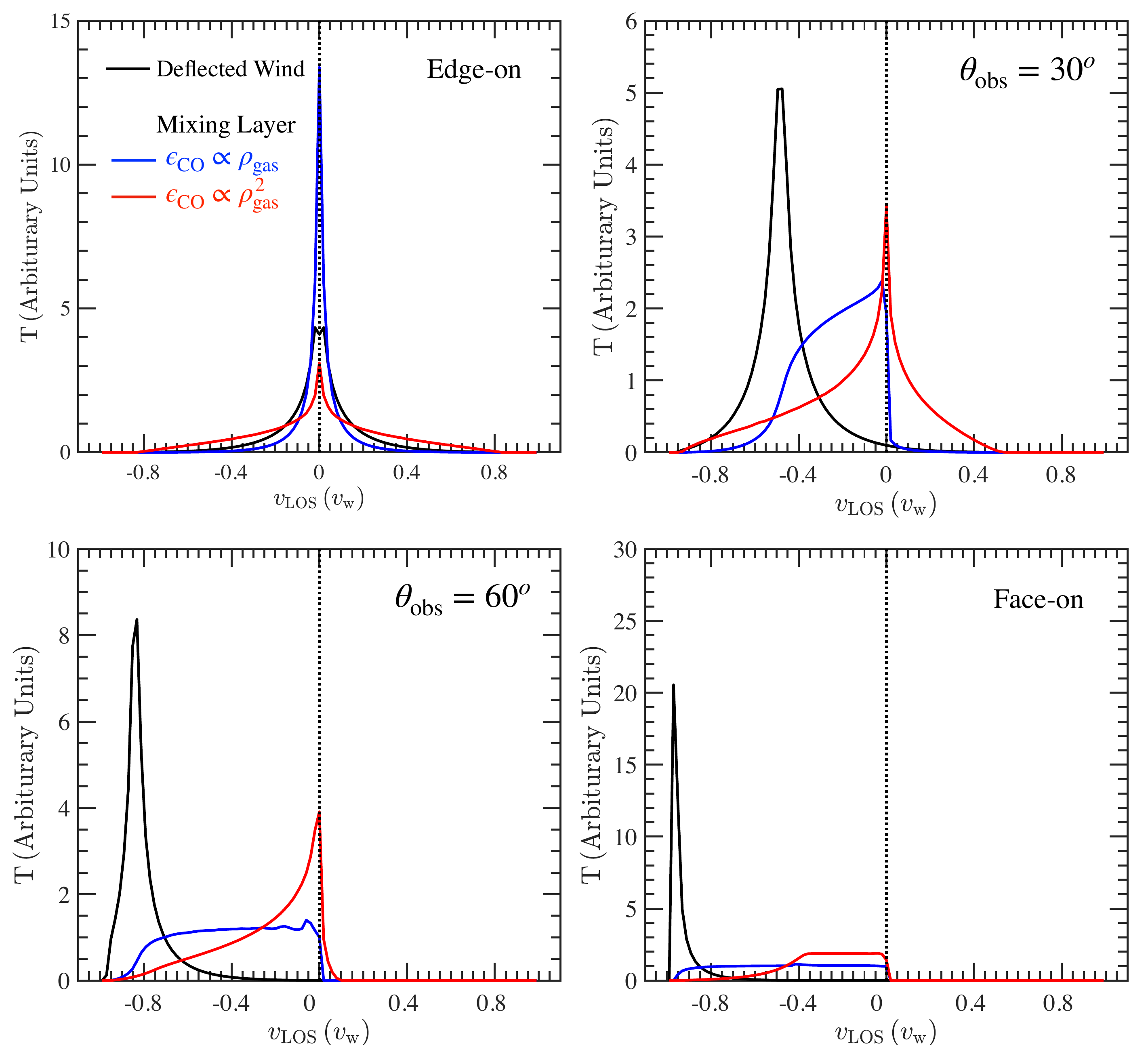}
	\end{center}
    \caption{Line profiles for the model with $\Lambda=25$ and $\lambda=1/2$. Each panel shows the result for a different direction of viewing as labeled, where $\theta_{\rm obs}$ is measured from the plane of the disk. Blue and red curves in each panel indicate the line profiles of material in the mixing-layer, under the assumption that emissivity of CO $J=$ 16 -- 15 ($\epsilon_{\rm CO}$) is proportional to the density and square of the density, respectively. For reference, we also show in each panel the emission profile of material in the shocked wind layer alone if no entrainment occurs, under the assumption that emissivity scales linearly to gas density (black curve).
    }
    \label{fig:lp}
\end{figure}

In Figure \ref{fig:lp} we show example line profiles for our reference model presented in the previous section ($\lambda$ = 1/2, $\Lambda = 25$).  We scale the projected velocities by $v_{\rm w}$ and integrate the emission up to $200\,r_{\rm d}$ ($0.6\,r_s$) from the base of the outflow. Four viewing angles are provided. The black curves show the line profiles for the shocked wind layer only, assuming emissivity proportional to density - mimicking the high density LTE regime. Except for the edge-on case, the emission always peaks near the projected wind velocity $v_{\rm w} \sin{\theta_{\rm obs}}$. This occurs because the bulk of the deflected wind flows at $v_1 \sim v_{\rm w}$ and is roughly parallel to the disk axis, due to the elongated shape of the cavity. The blue curves, on the other hand, show the computed line profiles for the mixing-layer only, assuming again that the emissivity is proportional to the density. As expected, the line profiles are much broader and flatter, peaking at zero and extending to a fraction of $v_{\rm w}$. Finally, the solid red curves show predicted line profiles from the mixing-layer when emissivity is proportional to the density squared, mimicking the low density limit. This emissivity condition increases the contribution of dense regions near the base, where the cavity opening angle is still large. Due to projection effects, it produces more extended line wings close to edge-on, and enhanced low-velocity emission when close to pole-on.




Our described model of partial entrainment of the wind, along with some of the exterior envelope, through a turbulent mixing-layer 
thus ensures that a fraction of the outflowing material is moving slowly due to the linear velocity gradient, Couette-type flow within the mixing-layer. Turbulent dissipation within the mixing-layer also provides a heating mechanism to make this material warmer than the shocked wind layer, hence brighter in high-$J$ CO lines.

In the following section, we investigate whether such a model could explain at the same time the line profile, momentum, and temperature of the broad component observed in CO by \textit{Herschel} towards the Serpens-Main SMM1 protostar, as well as the observed outflow cavity size, for reasonable envelope and wind parameters.


\section{Application to the broad component of Class 0 Protostar Serpens-Main SMM1}
\label{sec:AppC0}

As an illustration of the applicability of our formalism to a specific outflow case, we compare our model predictions with observations of the protostar Serpens-Main SMM1. This protostar is located in the Serpens Main cloud at a distance of 438 pc \citep{Herczeg19}. The protostar has a bolometric luminosity of $\sim$ 100 $L_\odot$ \citep{goicoechea12}, and is therefore on the border between low- and intermediate protostars. The envelope is correspondingly massive, with an estimated mass of $\sim$ 50 $M_\odot$ \citep{enoch08,Kristensen12}. When observed at high angular resolution, the protostar breaks up into multiple sources; however, the central most massive protostar is responsible for the primary outflow \citep{hull16, hull17, legouellec19}. When observed in H$_2$O and high-$J$ CO emission with the HIFI instrument on \textit{Herschel}, this source shows the brightest line intensity in H$_2$O and CO $J$=16--15 in the sample of \citet{Kristensen17}. For this reason, the broad line component of SMM1 also has the highest signal-to-noise ratio. Hence, this protostar is a natural choice for a first comparison between the model presented in this paper and observational data.

\subsection{Broad component line profile and wind velocity} 

When \textit{Herschel}-HIFI started observing H$_2$O emission toward protostars, one of the biggest surprises was that the velocity-resolved line profiles typically were dominated by a broad outflow component with a FWHM of $\gtrsim$ 30 km\,s$^{-1}$ \citep[e.g.,][]{Kristensen12, Kristensen17, mottram14}. This line width was significantly larger than seen in low-$J$ CO from the ground, e.g., $J$=3--2, where the FWHM is $\lesssim$ 15 km\,s$^{-1}$ \citep{Kristensen12}.  It also became clear that when observing higher-$J$ CO transitions with HIFI, the line profiles started resembling the H$_2$O profiles so much so that the CO $J$=16--15 profiles are indistinguishable from the H$_2$O profiles \citep{Kristensen17}. That the CO profiles vary with excitation suggests that the change in shape is indeed due to excitation as opposed to chemistry. Furthermore, the change in profile shape is likely related to an increase in temperature, because when calculating the rotational temperature from the ratio between CO lines, the temperature increases from $\lesssim$ 100 K to $\sim$ 300 K \citep{yildiz13, Kristensen17}. Thus, the higher-$J$ CO lines, and by implication the similar H$_2$O lines, trace a warmer, faster-moving, component of the protostellar outflow as compared to what is seen in low-$J$ transitions \citep{Kristensen17}, and this component is primarily seen as a broad outflow component in the velocity-resolved line profiles.

\citet{Kristensen12} and \citet{mottram14} speculated that, because of the higher temperature and velocity, this broad component is tracing gas closer to a shock front, 
possibly located where the protostellar wind shocks against the infalling envelope in an irradiated C-type shock. 
The colder gas, traced by lower-$J$ CO lines, then would be the subsequently entrained swept-up ambient gas. 
Alternatively, the heating and entrainment process of the broad component could take place within a turbulent mixing-layer at the interface between the shocked wind and infalling envelope. This alternate scenario is investigated below, using the model results presented in the previous sections.

\begin{figure}[ht]
    \begin{center}
	\includegraphics[width=0.6
	\columnwidth]{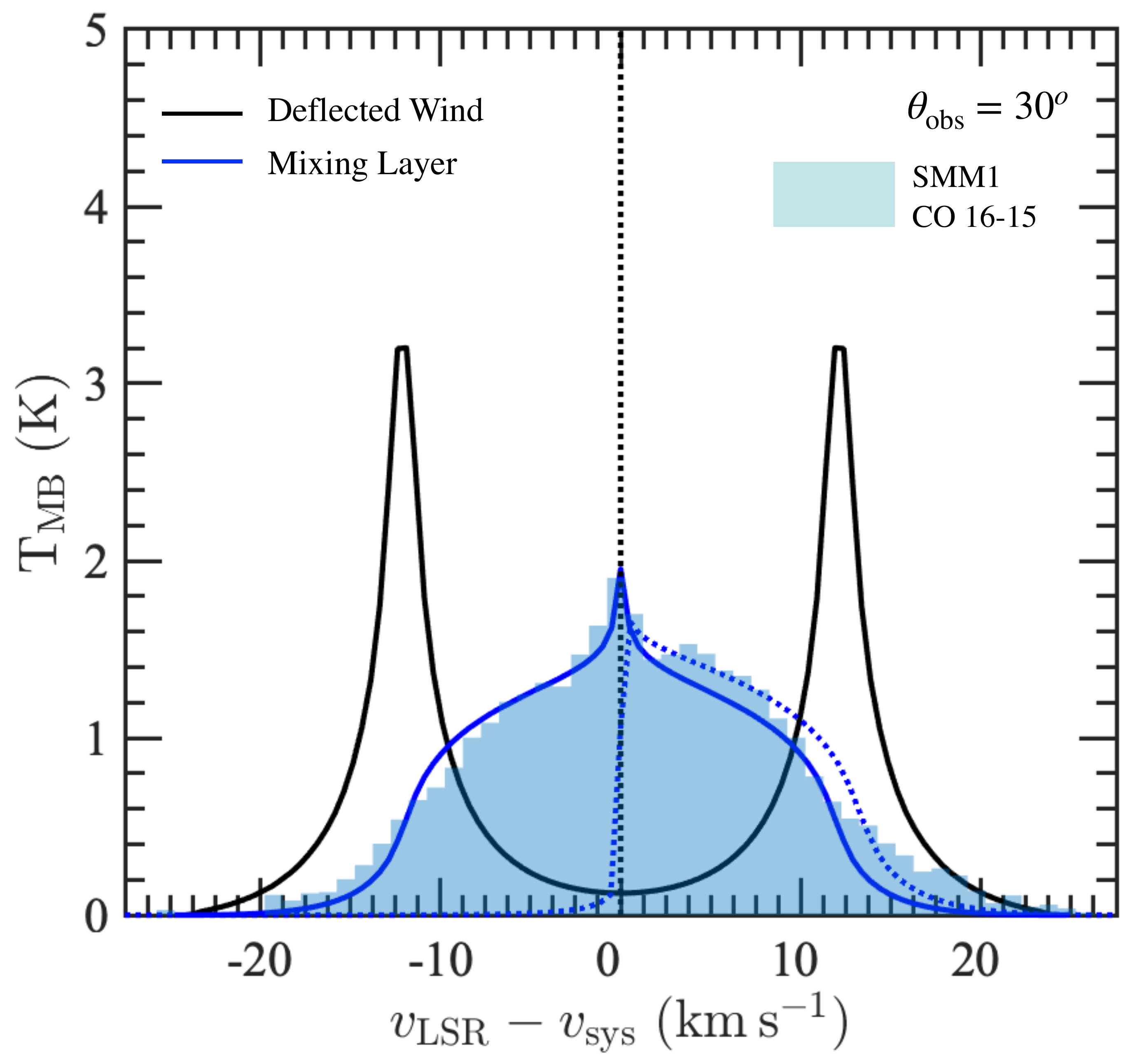}
	\end{center}
	\vspace{-10 pt}
    \caption{Observed broad component of the CO $J$=16--15 line profile in SMM1 (filled blue) compared against synthetic line profiles produced in Section \ref{sec:profile}. The synthetic profiles are generated at a viewing angle $\theta_{\rm obs} = 30\degr$ from the disk plane, with an emissivity that scales linearly to gas density. Blue and black curves assume that the CO emission originates from the central mixing-layer or from the deflected wind layer, respectively. The solid profiles include both outflow lobes and are re-scaled to $v_{\rm w}=25$ $\rm km\;s^{-1}$ so that the blue line can well match the left side of the observed SMM1 profile. An additional blue dotted fit is provided for the redshifted lobe (right side) of the SMM1 profile, requiring a somewhat larger $v_{\rm w}=28$ $\rm km\;s^{-1}$.}
    \label{fig:smm_line}
\end{figure}

For the source SMM1, \citet{Kristensen17} found that the CO $J$=16--15 line profile could be decomposed into three Gaussian components, one broad (FWHM $\sim$ 20 km s$^{-1}$) and two narrower components (FWHM $\sim$ 8--10 km s$^{-1}$). The narrower components are only seen in this high-$J$ CO line and likely originate in shocks very close to the protostar \citep{Kristensen13}, and they are not considered further here.
Figure \ref{fig:smm_line} compares the broad component, extracted from the CO $J$=16--15 line profile in SMM1 by \citet{Kristensen17} after removal of the two narrower component Gaussian fits, with our model predictions. Excellent agreement is found for a mixing-layer with $v_{\rm w} = 25-30$ km/s, an inclination of $\theta_{\rm obs}=30\degr$ measured from the plane of the disk, and an emissivity proportional to density.

A few checks are in order to ensure that the model remains self-consistent. First, for the ``thin shell" approximation to be satisfied, the wind velocity must remain large enough to produce a significant ram pressure which in turn confines the flow of shocked wind along the cavity surface. In Section \ref{sec:flow:nomix}, we showed that confinement requires $v_{\rm w}/c_w > 10$ where $c_w$ is the isothermal sound speed in the wind. With our estimate for $v_{\rm w} \simeq 30$ km\,s$^{-1}$, we therefore require $c_w < 3$ km\,s$^{-1}$, or a maximum wind temperature of $1000\,\mu$~K, with $\mu$ the mean molecular weight per particle (in a.m.u). This condition holds both in D-winds heated by ambipolar diffusion in Class 0 sources \citep{panoglou12,yvart16} and in X-winds in the absence of mechanical heating \citep{shang02}. In Section \ref{sec:flow:nomix}, we also found that the deflected envelope layer will always remain thin when $\Lambda > 10$, regardless of the value of $v_d$. We will verify that this condition holds in SMM1 in Section \ref{sec:disk}.

Second, the model line profiles in blue that reproduce the observed line profile shape for SMM1 in Figure \ref{fig:smm_line} are obtained only if a Couette linear velocity gradient exists across the mixing-layer. For this gradient to be maintained, the mass flow entrained in the mixing-layer from the wind side, $\dot{M}_{L1}$, should not exhaust the available flux of shocked wind material $\dot{M}_{1}$ flowing along the shell. As discussed in Section \ref{sec:flow:mix}, this constraint sets an upper limit on the turbulent entrainment coefficient in SMM1 $\alpha \le \alpha_{\rm max} = ({c_L}/{v_{\rm w}}) \simeq$ 0.035 (see Eqn.~\ref{eqn:alpha} and Figure \ref{fig:ml1}) where we have used our estimate of $v_{\rm w} =$ 25-30 km\,s$^{-1}$ from line profile modeling, and $c_L=$ 1 km\,$^{-1}$ from the temperature $\simeq 250$\,K inferred by multi-line CO analysis of the broad component in SMM1 \citep[see][and next section]{Kristensen17}. We will verify below that this upper limit on $\alpha$ is still compatible with the observed momentum in the broad component of SMM1.

\subsection{Outflow Cavity Size}
\label{sec:cavity-size}

\begin{figure*}[ht]
    \begin{center}
    \includegraphics[width=0.8 \columnwidth]{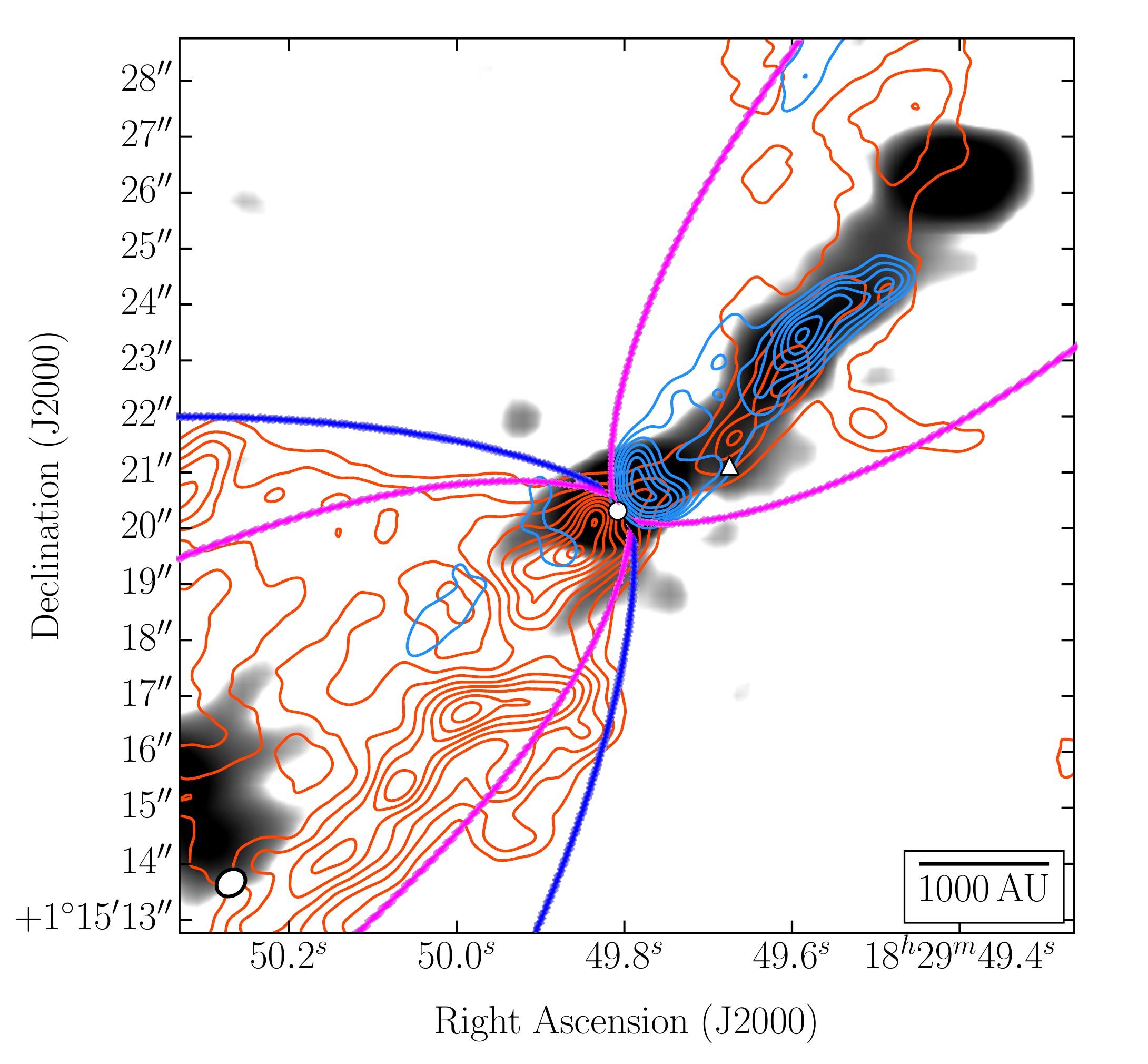}
    \vspace{-10 pt}
	\end{center}
    \caption{The model cavity shapes superimposed onto the observed CO map from \citet{hull16}. The magenta and blue lines correspond to $\Lambda=25$ and $\Lambda=50$ in our model, respectively (\cf~Figure~\ref{fig:shape}). They have been rescaled to $r_s=10,000$ au (magenta) and $r_s=40,000$ (blue) in order to match the observed CO map profile.
    }
    \label{fig:smm_shape}
\end{figure*}

In Figure \ref{fig:smm_shape} we compare a published CO outflow map of SMM1 \citep{hull16} with our predicted self-similar cavity shape from Figure \ref{fig:shape}, for various values of the scaling parameter $r_s$. Although only the inner region of the SMM1 outflow has been mapped at high angular resolution, the joint constraints on small and large scales indicate that $r_{\rm s}$ must lie in the range $\simeq 10,000-40,000$ au. Hence, we adopt $r_{\rm s} =$ 20,000 au as our fiducial value in the following. 

The cavity physical scale $r_s$ requires a specific ratio of
mass loss rate in the  wind to infall rate in the envelope (see Eqn. \ref{eqn:rs}),  
given by 
\begin{equation}
\frac{\dot M_{\rm w}}{\dot M_{\rm inf}} =  \frac{c_s^2}{v_{\rm w}} \, \left(\frac{r_s}{2GM_\star}\right)^{0.5}
\simeq 0.035\,
\left( \frac{c_s}{0.4\, {\rm km~s}^{-1}} \right)^2
\left( \frac{30\, {\rm km~s}^{-1}}{v_{\rm w}} \right)
\left( \frac{r_s}{20,000\, {\rm au}} \right)^{1/2}
\left( \frac{0.2\,M_\odot}{M_\star} \right)^{1/2}.
\label{eqn:mw2minf}
\end{equation}
Since SMM1 is quite bright (100 $L_\odot$) we adopt a fiducial sound speed in the envelope of $c_s \simeq 0.4$ km s$^{-1}$ corresponding to a temperature of 40~K. This value is consistent with radiative-transfer modeling of the dust emission from the envelope surrounding SMM1 \citep{Kristensen12}, which recover a sound speed 0.3 -- 0.5 km s$^{-1}$ in the envelope at the relevant physical scales, 350 -- 3000 au, shown in Fig. \ref{fig:smm_shape}, which are also those of the \textit{Herschel/HIFI} beam. With this value of $c_s$, we find that the observed size of the outflow cavity in SMM1 can be reproduced with a wide-angle wind mass-flux on the order of 4\% of the envelope infall rate, which is quite modest. In the following sections, we use our model and the observed momentum in the broad component to constrain the absolute value of $\dot M_{\rm w}$ and then that of $\dot M_{\rm inf}$, through Eqn.~\ref{eqn:mw2minf}.

\subsection{Momentum in the mixing-layer}
\label{sec:cavity-mom}

The two-sided momentum in the broad component of SMM1 was estimated from observations of CO $J$=3--2, 6--5, 10--9, and 16--15 taken with the JCMT, APEX, and \textit{Herschel}-HIFI \citep{yildiz13}. The respective line profiles were rebinned to the same velocity scale and to channels of 3 km\,s$^{-1}$ width. For each channel, a rotational diagram was constructed and, assuming LTE and optically thin emission, the rotational temperature and CO column density $N_{\rm CO}$ were calculated. The rotational temperature was $\sim$ 250 K, irrespective of velocity. With this mass spectrum in place, the mass-weighted momentum $\Pi_{\rm BC}$ of the broad component summed over both lobes is given by
\begin{equation}
    \Pi_{\rm BC} = \pi R_{\rm b}^2 \,
\left(\frac{1.4 m_{\rm H}}{X_{\rm CO}}\right) \, \int N_{\rm CO}(v)\, \mid v \mid \, dv \equiv \Pi_{\rm obs} \left(\frac{5 \times 10^{-5}}{X_{\rm CO}} \right),
\label{eqn:mom-obs}
\end{equation}
where $X_{\rm CO}$ is the (unknown) fractional abundance of CO molecules by number with respect to H nuclei in the broad component, and $\Pi_{\rm obs}$ is the fiducial ``observed" momentum assuming a standard interstellar CO abundance of $5\times 10^{-5}$. Equation \ref{eqn:mom-obs} shows that the value of $\Pi_{\rm obs}$ only depends on the observed CO intensity and excitation temperature, irrespective of the true $X_{\rm CO}$. It is therefore the quantity usually reported in observational papers. Using an updated distance, $d = 438\,$pc, to the Serpens Main cloud \citep{Herczeg19}, we recalculate the \citet{yildiz13} derived fiducial momentum inside a beam radius of $R_{\rm b} =$ $5.5^{\prime\prime}$ = 2400\,au to be $\Pi_{\rm obs} \simeq 8 \pm 2 \times 10^{-2} M_\odot$ km~s$^{-1}$. 

In our entrainment model, the two-sided momentum contained in the mixing-layer up to a distance $z = \pm R_{\rm b}$ is given by
\begin{equation}
\begin{split}
2 \Pi_L & = 2 \times \int_0^{R_{\rm b}} dz\, \int_0^h \rho_L(h,z) v_L(h,z) 2 \pi R(z)\, dh \\
    & = 2 \times \int_0^{R_{\rm b}} \dot{M}_L(z)\, dz \\
    & =  \dot{M}_{\rm w} \, \left(\frac{\alpha v_{\rm w}}{c_L}\right) 
    \times \int_0^{R_{\rm b}} \eta(z)\, dz 
    \simeq \dot{M}_{\rm w} \, \left(\frac{\alpha}{\alpha_{\rm max}} \right)
    \times \eta_{\rm b} R_{\rm b},
\end{split}
\label{eqn:momentum}
\end{equation}
where $\eta(z)$ is the normalized ratio plotted in Figure \ref{fig:mlt} and $\alpha_{\rm max}$ is defined in Eqn.~\ref{eqn:alpha}. Since  $\eta(z)$ increases very slowly with height, the integral on $z$ may be approximated as $\eta_{\rm b} R_{\rm b}$, where $\eta_{\rm b} \equiv \eta(R_{\rm b})$. 

If the  momentum  in the broad component of SMM1 inside $R_{\rm b}$ = 2400 au is provided by mixing-layer entrainment from a wide-angle wind, then $2\Pi_L = \Pi_{\rm BC}$. 
Using Equation \ref{eqn:mom-obs} with $\Pi_{\rm obs} \simeq 8 \pm 2 \times 10^{-2} M_\odot$ km~s$^{-1}$, and taking $\eta_{\rm b} \simeq 1$ (which we will verify in Section \ref{sec:disk}) we infer that the wide-angle wind must have a mass flux 
\begin{equation}
\begin{split}
\dot{M}_{\rm w} & = \left( \frac{\alpha_{\rm max}}{\alpha} \right) \left(\frac{\Pi_{\rm BC}}{\eta_{\rm b}R_{\rm b}}\right) \\
 & \simeq 6 \times 10^{-6}\ \left( \frac{\alpha_{\rm max}}{\alpha} \right) 
 \left( \frac{5 \times 10^{-5}}{X_{\rm CO}} \right) 
 M_\odot\, {\rm yr}^{-1}.
\end{split}
\label{eqn:mwdotobs}
\end{equation}

\subsection{Infall rate}
\label{sec:cavity}

We can now compute the required envelope infall rate in our model, and compare with the infall rate independently suggested by dust envelope models. 
Combining Equations \ref{eqn:mw2minf} and \ref{eqn:mwdotobs}, 
we infer the required infall mass flux to reproduce both the outflow cavity size and 
the momentum in the broad component as
\begin{equation}
\begin{split}
\dot{M}_{\rm inf} & 
= \dot{M}_{\rm w} \left( \frac{\dot M_{\rm inf}}{\dot M_{\rm w}} \right) \\
 & \simeq 1.4 \times 10^{-4}\, M_\odot \, {\rm yr}^{-1} 
 \left( \frac{\alpha_{\rm max}}{\alpha} \right)
 \left( \frac{5 \times 10^{-5}}{X_{\rm CO}} \right)
 \left( \frac{M_\star}{0.2M_\odot} \right)^{1/2}.
\end{split}
\label{eqn:mdotinf}
\end{equation}
The required value is larger than typical infall rates for low-mass Class 0 protostars. It is in line, however, with that expected for sources with particularly massive envelopes such as SMM1, 
whose luminosity $\sim$ 100 $L_\odot$ places
it on the border between low- and intermediate-mass protostars. 
Using the estimated H$_2$ density at 1000~au, $n_{1000}$, in the SMM1 dust envelope model of \citet{Kristensen12} and rescaling by $d^2$ from $d=$ 230~pc to 438 pc, we infer an ``observed" envelope infall rate at $R_{1000}$ = 1000 au of
\begin{equation}
\begin{split}
\dot{M}_{\rm env} &= 4 \pi R_{1000}^2\, n_{1000}\, v_{\rm inf}(R_{1000}) \times (1.4m_{\rm H_2}),\\
& \simeq 1.7 \times 10^{-4}\ M_\odot \, {\rm yr}^{-1}  \left( \frac{n_{1000}}{1.5 \times 10^7 {\rm cm}^{-3}} \right)
            \left( \frac{M_\star}{0.2M_\odot} \right)^{1/2},
\end{split}
\label{eqn:mdotenv}
\end{equation}
where the factor 1.4 accounts for the mass in the form of Helium. We note that $M_\star$ appears at the same power in $\dot{M}_{\rm inf}$ and $\dot{M}_{\rm env}$, hence its exact value, currently unknown in SMM1, does not matter for the comparison. There is therefore good agreement with our mixing-layer model as long as $X_{\rm CO}$, the CO abundance in the mixing-layer with respect to H nuclei, is close to the standard interstellar value of $5\times 10^{-5}$, and the turbulent entrainment parameter $\alpha$ is close to the maximum value to maintain a Couette flow, $\alpha_{\rm max} = c_L / v_{\rm w} \simeq$  0.03. We note that such a value of $\alpha$  matches very well with a model fit to supersonic mixing-layer experiments\footnote{\citet{cr91} showed that the variation of opening angle versus Mach angle in experiments could be reproduced with $\epsilon \equiv v_{\rm ent}/c_2 = 0.089\epsilon_2$ with $\epsilon_2 \equiv c_2/(3c_L)$; this is equivalent to our adopted prescription $v_{\rm ent} = \alpha c_2^2/c_L$ with $\alpha \simeq 0.03$.}, thus it appears physically plausible. With these values for $X_{\rm CO}$ and $\alpha$, the wind mass-flux that is required to provide the broad component momentum is $\dot M_{\rm w} \simeq 6 \times 10^{-6} M_\odot\, {\rm yr}^{-1}$ (see Eqn.~\ref{eqn:mwdotobs}).



\subsection{Constraints on ejection / accretion ratio and disk radius}
\label{sec:disk}
We have shown in Section~\ref{sec:cavity-size} that the outflow cavity size in SMM1 can be reproduced with a modest ratio of wind mass flux to envelope infall rate of $\simeq 4$\%. 
The disk accretion rate onto the central star, however, may be smaller than the envelope infall rate onto the disk. Assuming that the bolometric luminosity $\simeq$ 100$L_\odot$ of the SMM1 source \citep{goicoechea12} is dominated by the accretion luminosity $L_{\rm acc} \simeq GM_\star \dot M_{\rm acc}/R_\star$, and adopting stellar radii $R_\star$ on the birthline computed by \citet{Hosokawa09}, we infer a disk accretion rate of $\dot{M}_{\rm acc} \simeq 10^{-4}\, M_\odot$\,yr$^{-1}$ if $M_\star = 0.2\,M_\odot$, and $\dot{M}_{\rm acc} \simeq 3 \times 10^{-5}\, M_\odot$\,yr$^{-1}$ if $M_\star = 0.5\,M_\odot$.
With the wide-angle wind mass-flux $\dot M_{\rm w} \simeq 6 \times 10^{-6} M_\odot\, {\rm yr}^{-1}$
derived in the previous section, the ratio of wind ejection rate to disk accretion rate is thus $\simeq 0.06-0.2$ for $M_\star = 0.2-0.5 M_\odot$. Such values are in the typical range predicted by D-wind and X-wind ejection models from accretion disks around young stars.
Therefore, the wind mass-flux requirements in our model for SMM1 appear physically reasonable.

We  next estimate the expected range of the parameter $\lambda$, the ratio of wind to infall ram pressure, for SMM1. From Eqn.~\ref{eqn:lambda} and Eqn.~\ref{eqn:rs} we have
\begin{equation}
\begin{split}
\lambda & = \frac{v_{\rm w}\, \dot{M}_{\rm w}}{v_{\rm d}\,\dot{M}_{\rm inf}} 
= \left(\frac{c_s^2}{GM_\star}\right) \sqrt{\frac{r_{\rm s} \times r_{\rm d}}{2}} \\ 
& \simeq 0.5 \left( \frac{r_{\rm s}}{20,000\, {\rm au}} \right)^{1/2}
\left( \frac{r_{\rm d}}{30\, {\rm au}} \right)^{1/2}
\left( \frac{c_s}{0.4\, {\rm km~s}^{-1}} \right)^2
\left( \frac{0.2\,M_\odot}{M_\star} \right).
\end{split}
\end{equation}
This is consistent with the condition $\lambda \gtrsim 0.2$ for which our cavity solutions break out and reach their full extent (see Figure \ref{fig:width}), for the typical disk sizes in Class 0 sources \citep{Maury19}. Interferometric continuum observations suggest that SMM1 possesses a particularly large and massive disk of $\simeq 300$ au \citep{Enoch09}, hence the breakout condition is very likely fulfilled.

Finally, we estimate the typical $\Lambda$ parameter as
\begin{equation}
\Lambda  = \sqrt{\frac{2 r_{\rm s}}{r_d}}
 = 36 \times \left( \frac{r_{\rm s}}{20,000\, {\rm au}} \right)^{1/2}
\left( \frac{30\, {\rm au}}{r_{\rm d}}\right)^{1/2}.
\end{equation}
This low value of $\Lambda$ is consistent with our assumption of $\eta_{\rm b} \simeq 1$ on the scale $z = R_{\rm b} \simeq 0.1r_s$ of the \textit{Herschel} beam (cf. the curves for $\eta(z)$ in the right panel of Figure \ref{fig:mlt}). Even with a large disk, $r_d \simeq 300$ au, the condition $\Lambda > 10$ for the weakly shocked deflected envelope layer to remain thin is also fulfilled.

\subsection{Temperature and Density in the mixing-layer}
\label{sec:temperature}

As an additional test of our model, we investigate whether the observed temperature,  $T_L \simeq$ 250 K of the broad component in SMM1 suggested by multi-line CO analysis \citep{Kristensen17} would be consistent with heating of the mixing-layer in our model by turbulent viscosity. 

In principle, a full non-equilibrium thermo-chemical calculation should be performed as a function of position along the mixing-layer. Such a complex problem is, however, outside the scope of the present paper and is deferred to future work. For simplicity, we assume here that the temperature and chemistry in the mixing-layer have reached a steady-state on the scales observed by the \textit{Herschel} HIFI beam, and check whether thermal equilibrium at $T_L \simeq$ 250 K could indeed be sustained.

Following \citet{Binette99}, we take a turbulent viscosity $\mu = (\alpha/4) \rho_L c_L h$ with $h$ the total thickness of the mixing-layer. A derivation of this expression for supersonic isothermal mixing-layers with a linear velocity profile (Couette flow) is given  in Appendix C. We can then express the turbulent heating rate per unit volume in the mixing-layer as
\begin{equation}
\begin{split}
\Gamma_{\rm visc} & = \mu \,\left( \frac{dv}{dh} \right)^2
= \left( \frac{\alpha \rho_L c_L}{4h}\right) \,  v_{\rm w}^2 \\
& = \left( \frac{\alpha v_{\rm w}}{c_L}\right) \left(\frac{P v_{\rm w}}{4h}\right) \\
& = \left( \frac{\alpha/\alpha_{\rm max}}{2 \dot M_L} \right)\, \pi R\, P\, \rho_L\, \overline{v_L} v_{\rm w},
\label{eqn:gamma}
\end{split}
\end{equation}
where we make use of $(dv/dh) = v_{\rm w}/h$ with $h = \dot M_L / (2\pi R \rho_L \overline{v_L})$, $R$ is the local shell radius, 
$P = \rho_L c_L^2$ is the local pressure in the layer, and $\alpha_{\rm max} = c_L / v_{\rm w}$ is defined in Eqn.~\ref{eqn:alpha}.

Thermal equilibrium at constant $T_L$ will be maintained as long as 
\begin{equation}
\Gamma_{\rm visc} \simeq \Lambda_{\rm exp} + \Lambda_{\rm rad},
\end{equation}
where $\Lambda_{\rm exp}$ is the rate of ``expansion cooling" in the mixing-layer as the pressure $P$ drops with altitude and $\Lambda_{\rm rad}$ is the radiative cooling rate (both per unit volume). The contribution of H$_2$ formation is neglected in this analysis, as well as advection of thermal energy into the layer, since $c_1 \ll v_{\rm w}$ and $c_2 \ll v_{\rm w}$. Under our isothermal hypothesis for the mixing-layer, the expansion cooling rate may be simply expressed as
\begin{equation}
\Lambda_{\rm exp} = - \frac{dP}{dt} = \overline{v_L}\, \left(\frac{-dP}{dx}\right)  = \overline{v_L} \left( \frac{P}{x}\right)  
\left( \frac{-d\log P}{d\log x}\right),
\label{eqn:lambda_exp}
\end{equation}
where $x$ denotes position along the layer. We thus obtain 
\begin{equation}
\begin{split}
\frac{\Lambda_{\rm exp}}{\Gamma_{\rm visc}} & = 
\left( \frac{\alpha_{\rm max}}{\alpha} \right) \frac{4h}{x} \left( \frac{\overline{v_L}}{v_{\rm w}} \right) \left(\frac{-d\log P}{d\log x} \right)\\
& = 2\dot M_{\rm L}\, \left(\frac{\alpha_{\rm max} }{\alpha} \right) \left(\frac{1}{\pi R x \rho_L v_{\rm w}}\right)
\left(\frac{-d\log P}{d\log x} \right)\\
& =  \eta(z)\, \left(\frac{\dot M_{\rm w}\,v_{\rm w}}{P(z)\,4 \pi r_s^2}\right)  \left( \frac{4 r_s^2}{R x} \right) \left( \frac{c_L}{v_{\rm w}} \right)^2 \left(\frac{-d\log P}{d\log x}\right),
\end{split}
\label{eqn:exp2visc}
\end{equation}
where $\eta$ is the normalized mass flux within the mixing-layer plotted in Figure \ref{fig:mlt}. Therefore, in our model, the ratio of expansion cooling to viscous heating in the mixing-layer is independent of the turbulent entrainment efficiency $\alpha$, and only scales with $({c_L}/{v_{\rm w}})^2$. Furthermore, the remaining terms in this ratio are only weakly dependent on the values of $\lambda$ and $\Lambda$ (see Figures \ref{fig:npml} and \ref{fig:mlt}). On the typical scale $z \leq 0.1\,r_s$ encompassed by the \textit{Herschel}/HIFI beam, we find that $(\Lambda_{\rm exp}/\Gamma_{\rm visc})$ $\le 200\,(c_L / v_{\rm w})^2$. With our fitted values of $c_L \simeq 1$ km/s and $v_{\rm w} \simeq$ 30 km/s for the broad component of SMM1, we infer that expansion cooling should be negligible with respect to viscous heating.

Thus, we only need to compare the viscous heating rate with the radiative cooling rate. As noted by \citet[][see their Figure~11]{Kristensen17}, cooling by CO largely dominates over cooling by H$_2$ at temperatures of 250~K (for a standard CO/H$_2$ abundance ratio). We further assume that CO cooling is excited mainly by collisions with H$_2$ in the low-density limit (which we will verify a posteriori for SMM1). Denoting $L_0(T)$ as the CO cooling rate coefficient (in ${\rm erg\,s}^{-1}$\,cm$^{3}$) at temperature $T$, and $X_{\rm CO}$ and $X_{\rm H_2}$ as the CO and H$_2$ abundances relative to the total number density of H nuclei, $n_{\rm H} = \rho_L / (1.4\, m_{\rm H})$, we have
\begin{equation}
\Lambda_{\rm CO} = L_0(T_L) \, n({\rm CO}) \, n({\rm H_2}) =  L_0(T_L) \, X_{\rm CO} \, X_{\rm H_2} \, \left ( \frac{\rho_L}{1.4m_H} \right)^2.
\label{eqn:cooling}
\end{equation}
The ratio of turbulent heating to CO cooling is then independent of the mixing-layer density $\rho_L$. With a mean layer velocity $\overline{v_L} \simeq v_{\rm w}/2$ (Couette flow), a typical value of $2\dot M_L$ within the \textit{Herschel} beam of $2\Pi_L / R_{\rm b}$
(see Eqn.~\ref{eqn:momentum}), and $2\Pi_L = \Pi_{\rm BC}$ where $\Pi_{\rm BC}$ is the momentum in the warm CO broad component, this ratio can be expressed as
\begin{equation}
\begin{split}
\frac{\Gamma_{\rm visc}}{\Lambda_{\rm CO}}  &= \left( \frac{c_L^2}{L_0(T_L) X_{\rm CO} X_{\rm H_2}} \right)
\left( \frac{\pi R R_{\rm b} v_{\rm w}^2}{2 \Pi_{\rm BC}} \right)\,
\left( \frac{\alpha}{\alpha_{\rm max}} \right)\,
(1.4\, m_{\rm H})^2 \\
&\simeq 0.4 \left( \frac{0.5}{X_{\rm H_2}} \right) \left( \frac{c_L}{1\, {\rm km\,s}^{-1}} \right)^2 
\left( \frac{3 \times 10^{-24}\, {\rm erg\,s^{-1}\, cm^{3} } }{L_0(T_L)} \right)
\left( \frac{v_{\rm w}}{30\, {\rm km\, s}^{-1}} \right)^2
\left( \frac{R_{\rm b}}{2400\, {\rm au}} \right)\\
& 
\left( \frac{R}{1000\, {\rm au}} \right) 
\left( \frac{\alpha}{\alpha_{\rm max}} \right)
\left( \frac{8 \times 10^{-2}\, M_\odot {\rm km\, s}^{-1}}{\Pi_{\rm obs}} \right).
\end{split}
\label{eqn:heatcoolrat}
\end{equation}
It is remarkable that apart from $X_{\rm H_2}$ there are no free parameters in this ratio, as all of the other factors are well constrained by observations of SMM1: The value of $c_L$ is fixed by the relative intensities of the high-$J$ CO lines, indicating
$T_L \simeq 250$ K. The corresponding value of $L_0 \sim 3 \times 10^{-24}\, {\rm erg}\,$s$^{-1}$\,cm$^{3}$ at 250~K is set by molecular collision rate calculations \citep{neufeld93}. The value of $v_{\rm w}$ derives from our model fit to the CO(16-15) line profile in Figure \ref{fig:smm_line}. The cavity radius $R \simeq 1000$ au at $z = R_b =$ 2400 au derives from our model fitting of the outflow shape in Figure \ref{fig:smm_shape}. The value of $\alpha \simeq \alpha_{\rm max}$ is required for our model to be consistent with the dust envelope infall rate in SMM1 (see Section \ref{sec:cavity}).
Finally, the product $\Pi_{\rm BC} \,X_{\rm CO}$ is equal to $\Pi_{\rm obs} \times (5 \times 10^{-5})$, where the value of
$\Pi_{\rm obs} = 8 \times 10^{-2}\, M_\odot\, {\rm km\, s}^{-1}$ is fixed by the observed CO line profile intensity and excitation temperature in SMM1 (see Equation \ref{eqn:mom-obs}).

We conclude that if hydrogen is mostly in molecular form ($X_{\rm H_2} \simeq 0.5$), and CO cooling is not far from the low-density regime, the ratio in Eqn.~\ref{eqn:heatcoolrat} is close to 1 for our mixing-layer model of SMM1, and thermal equilibrium can be maintained at the observed temperature $\simeq 250$ K of the broad CO $J$=16--15 component.

The low-density CO cooling expression applies only until $L_0\, n({\rm H_2}) \simeq 0.5\, \mathcal{L}_{LTE}$, where $\mathcal{L}_{LTE}$  is the cooling rate per CO molecule in the high-density LTE regime. At 250 K, $\mathcal{L}_{LTE} \simeq 10^{-18}$ erg/s \citep{neufeld93} hence the validity extends to $n({\rm H_2}) \leq 1.7 \times 10^5$ cm$^{-3}$. To estimate the density in the SMM1 mixing-layer on the scale of the HIFI beam, we note that the shell pressure distribution on large scales is approximated by Eqn.~\ref{eqn:pressure}. We infer the H nucleus density predicted in the mixing-layer at $Z \simeq R_b$ for the SMM1 model parameters
\begin{equation}
\begin{split}
n_{\rm H} &= \frac{P(R_b)}{1.4\,m_H c_L^2},\\ 
&\simeq 1.4 \times 10^5\, {\rm cm}^{-3} \left( \frac{2400\, {\rm au}}{R_{\rm b}} \right)^{1.5}\, 
\left(\frac{\dot{M}_{\rm w}}{6 \times 10^{-6}\ M_\odot\, {\rm yr}^{-1}}\right)
\left(\frac{v_{\rm w}}{30\, {\rm km\,s}^{-1}}\right)
\left( \frac{20,000\, {\rm au}}{r_s} \right)^{0.5}
\left( \frac{1\, {\rm km\, s}^{-1}}{c_L} \right)^2.
\end{split}
\end{equation}
If all hydrogen is in molecular form, we have $n({\rm H_2}) = 0.5n_H \simeq 0.7 \times 10^5$ cm$^{-3}$  and the low-density regime of CO cooling assumed in Eqn.~\ref{eqn:heatcoolrat} is indeed justified for SMM1 on HIFI beam scales.

\subsection{Summary and Discussion of the Model Fit to Protostar Serpens-Main SMM1}
\label{sec:disc}


In summary, we have shown that our simple model of a turbulent mixing-layer across a static wind/envelope interface is able to reproduce successfully all of the observed properties of the broad CO outflow component discovered by \textit{Herschel}/HIFI in the Serpens-Main SMM1 protostar, 
for a self-consistent and physically realistic set of parameters. The CO $J$=16--15 line profile shape and velocity extent are reproduced for a typical wind speed $v_{\rm w} \simeq$ 30 km s$^{-1}$ and a view angle of $\theta_{\rm obs}=30\degr$ to the disk plane ({\ie}~the median value expected for random inclinations). This wind speed is smaller than predicted for an X-wind from the innermost disk radius at $\simeq 0.1$ au 
\citep[$v_{w} \simeq 150$ km/s;  see e.g.][]{shang98} 
but remains compatible with a slow MHD disk wind launched from a few au in the disk \citep[see e.g.][]{tabone20}.
Next, the observed outflow cavity size on $300-3000$ au scales, when combined with the estimated dust temperature in the envelope, requires a ratio of wind mass-flux to infall rate of 4\%. With this imposed ratio, the observed CO-emitting momentum in the broad component (provided by wind entrainment) is consistent with the observed infall rate in the dusty envelope for a standard interstellar CO abundance and a turbulent entrainment coefficient $\alpha \simeq$ 0.03 (consistent both with our assumption of a Couette flow in the mixing-layer, and with mixing-layer laboratory experiments). The corresponding wind mass-flux then represents a fraction $\simeq 0.06-0.2$ of the disk accretion rate onto SMM1 (as determined from its bolometric luminosity), consistent with current disk wind models. Finally, the observed temperature in the broad CO outflow component of SMM1 is consistent with a balance between turbulent heating and CO cooling in the mixing-layer if H$_2$ is mostly in molecular form, which is very likely at such low temperatures. We also verify that the values of $\lambda$ and $\Lambda$ in our SMM1 model are consistent with the conditions for cavity breakout and the requirement of thin shells across the full range of disk radii expected in such a source, $r_d =$ 10--300 au. 


A obvious next step for this modeling work would be to compute self-consistently the time-dependent evolution of temperature and chemistry through the wind shock and along the mixing-layer, using for example the molecular MHD disk wind models of \citet{panoglou12} and \citet{yvart16} as initial conditions. Such a calculation would provide an important check on our model requirement of an interstellar CO abundance in the mixing-layer of SMM1, to match independent constraints on the infall rate obtained from dust emission observations. It would also aid in the identification of the best tracer for the predicted narrow emission from the shocked wind layer (cf. black double-horned profile in Fig. \ref{fig:smm_line}).

Furthermore, since our model assumes a steady wind-blown cavity, it provides a natural explanation not only for the broad \textit{Herschel}-bright CO component, but also for the narrow outflow cavity radii $\leq 3000$ au observed at the Class 0 stage of $10^4-10^5$ yrs, despite observed CO velocities on the order of 10 km/s. In contrast, for wind-driven shell models with full mixing, quasi radial outflow motions of the same amplitude lead to excessive cavity radii in only a few thousand years \citep[see eg.][]{shang06,Lopez19}. Comparison over a larger sample of protostars with well characterized broad components and outflow cavities will be necessary to verify that self-consistent models can be found, as in SMM1, and to investigate how the required wide-angle wind properties would need to vary with source properties. 



\section{Conclusions}
\label{sec:conc}

In this paper we have reconsidered the interaction of a wind expanding into a surrounding medium under the assumption of partial mixing across the boundary layer separating the shocked wind and envelope. Our solutions differ from conventional wind/envelope interaction models where instantaneous full mixing is assumed \citep[eg.][]{li96,lee00,Lopez19} in that we produce static, rather than expanding, shells. To maintain the stationary shape, we allow the shocked and deflected wind to flow upward at close to $v_{\rm w}$ along the interior of the cavity wall while the shocked and deflected envelope moves slowly downwards along the exterior of the cavity wall. A turbulent entrainment layer is thus able to form between these two deflected flows. 

Specifically, we determine the shape of the stationary cavity  formed when an isotropic wind interacts with an infalling and rotating \citep{ulrich76} envelope. The resulting model is then quantitatively compared with observations of the protostellar outflow from SMM1 in the Serpens Molecular Cloud. 

The main results of our analysis are as follows:

\begin{enumerate}

\item The shape of the steady-state cavity (Section \ref{sec:shape:values}) is determined by two non-dimensional parameters, $\lambda$, the ratio of the wind ram pressure to the fiducial infall ram pressure (Eqn.\ \ref{eqn:lambda}), and $\Lambda$, the ratio of the wind ram pressure to the envelope thermal pressure at the edge of the disk (Eqn.\ \ref{eqn:Lambda}) . We show that $\Lambda$ sets the  foot-point of the cavity at the disk plane (Figure \ref{fig:base}) and that breakout solutions require $\lambda > 0.2$, with the cavity shapes becoming self-similar for $\lambda > 0.5$ (Figure \ref{fig:profile}).  In the self-similar regime, the size scaling of the cavity is determined by $r_s$ (Eqn.\ \ref{eqn:rs}).

\item Under the assumption of no mixing
(Section \ref{sec:flow:nomix}), the shocked and deflected wind moves upward along the cavity at close to the velocity $v_{\rm w}$, while the shocked and deflected envelope moves downward only slowly, except very near the base (Figure \ref{fig:mom}). Furthermore, away from the base the associated downward momentum flux is much less than the upward momentum flux (Figure \ref{fig:vel}). 

\item Under the assumption of partial mixing within a turbulent layer between the upward and downward shocked deflected layers (Section \ref{sec:flow:mix}), the overall amount of material brought into mixing-layer, from both sides, is directly proportional to the mass-loss rate in the wind multiplied by the entrainment efficiency $\alpha$ and $v_{\rm w}/c_L$ (Figure \ref{fig:mlt}). Furthermore, as previously shown by \citet{cr91}, the mass entrainment from the upward, wind, side is roughly twice that of the downward, envelope, side (Figure \ref{fig:ratio}), where the approximate proportionality is set by the  assumption that across the mixing-layer the flow velocity profile is linear ({\ie}~a Couette flow).

\item The shape of the line profile produced by material flowing along the cavity wall strongly depends on which layer is responsible for the emission (Section \ref{sec:line_profile}). The upward, shocked wind layer moves fast, $v \sim v_{\rm w}$, and has little curvature, resulting in a narrow profile peaked at the projected wind velocity. Alternatively, due to the Couette-type flow, emission 
from the mixing-layer is broad and peaks at rest velocity (Figures \ref{fig:lp} and \ref{fig:smm_line}).

\item We find an excellent correspondence between the broad component of the CO $J$=16-15 line profile observed by \textit{Herschel} towards the protostar Serpens-Main SMM1, and a mixing-layer model with $v_{\rm w} =$ 25--30 km/s, a viewing angle 30 degrees from the disk plane, and an emissivity proportional to density (Section \ref{sec:AppC0}). Furthermore, taking $\alpha \simeq 0.03$, a value which matches very well with experimental measurements of supersonic mixing, and assuming a standard CO abundance and a reasonable ratio of wind to infall rate of 4\%, we find excellent quantitative agreement between the observed momentum in the CO broad component, the observed infall rate of SMM1, and the observed outflow cavity size (Section \ref{sec:cavity}). 

\item We compute the turbulent heating, expansion cooling, and radiative CO cooling within the mixing-layer, and show that their ratio is appropriate to keep the gas warm at the observed temperature $T_L \simeq 250$ K in SMM1 (Section \ref{sec:temperature}).

\item Finally, our model provides a natural explanation for the narrow outflow cavity radii observed at the Class 0 stage of $10^4$ -- $10^5$\,yrs (Section \ref{sec:disc}). Unlike wind-driven shell models with full-mixing, in which radial motions quickly lead to large cavity sizes, our partial mixing solutions with a mixing layer separating the shocked wind and envelope 
produce a time-independent, steady cavity where observed velocities are parallel to the cavity walls, and thus do not lead to excessive expansion.

\end{enumerate}

To summarize, we provide a model for the interaction between a wind and a surrounding envelope which potentially can be applied widely, from protostellar outflows to galactic-scale. 
The model produces steady-state cavities, broad line profiles peaked at the rest velocity, and constrains the turbulent entrainment efficiency.
It therefore provides a new framework in which to interpret the observations of warm wind-driven outflows, and in particular to reconcile modest outflow cavity widths with the large observed flow velocities.
While the model successfully reproduces a number of observational constraints for a single protostellar outflow, Serpens-Main SMM1, an obvious next step is to apply this analysis to a larger sample of protostellar sources in order to test its success; this will be done in a forthcoming publication. 

\acknowledgements

We thank the anonymous referee for comments that improved this paper. DJ is supported by the National Research Council of Canada and by a Natural Sciences and Engineering Research Council of Canada (NSERC) Discovery Grant. The work of SC is supported by the Programme National Physique et Chimie du Milieu Interstellaire (PCMI) of CNRS/INSU with INC/INP and co-funded by CNES, and by the Conseil Scientifique of Observatoire de Paris. 
The research of LEK is supported by a research grant (19127) from VILLUM FONDEN. Research at the Centre for Star and Planet Formation is funded by the Danish National Research Foundation. 

DJ would like to thank the Observatoire de Paris, the Observatoire de Grenoble, and the European Southern Observatory in Garching, Germany for hosting him during visits at which many detailed discussions for this paper took place. 
DJ and SC also thank the \textit{Herschel} WISH team, especially Ewine van Dishoeck, for strongly encouraging this research project.

\appendix

\section{The boundary condition of the interface at the disK plane}
\label{ap:base}

\begin{figure*}[ht]
    \begin{center}
    \includegraphics[width=0.6 \columnwidth]{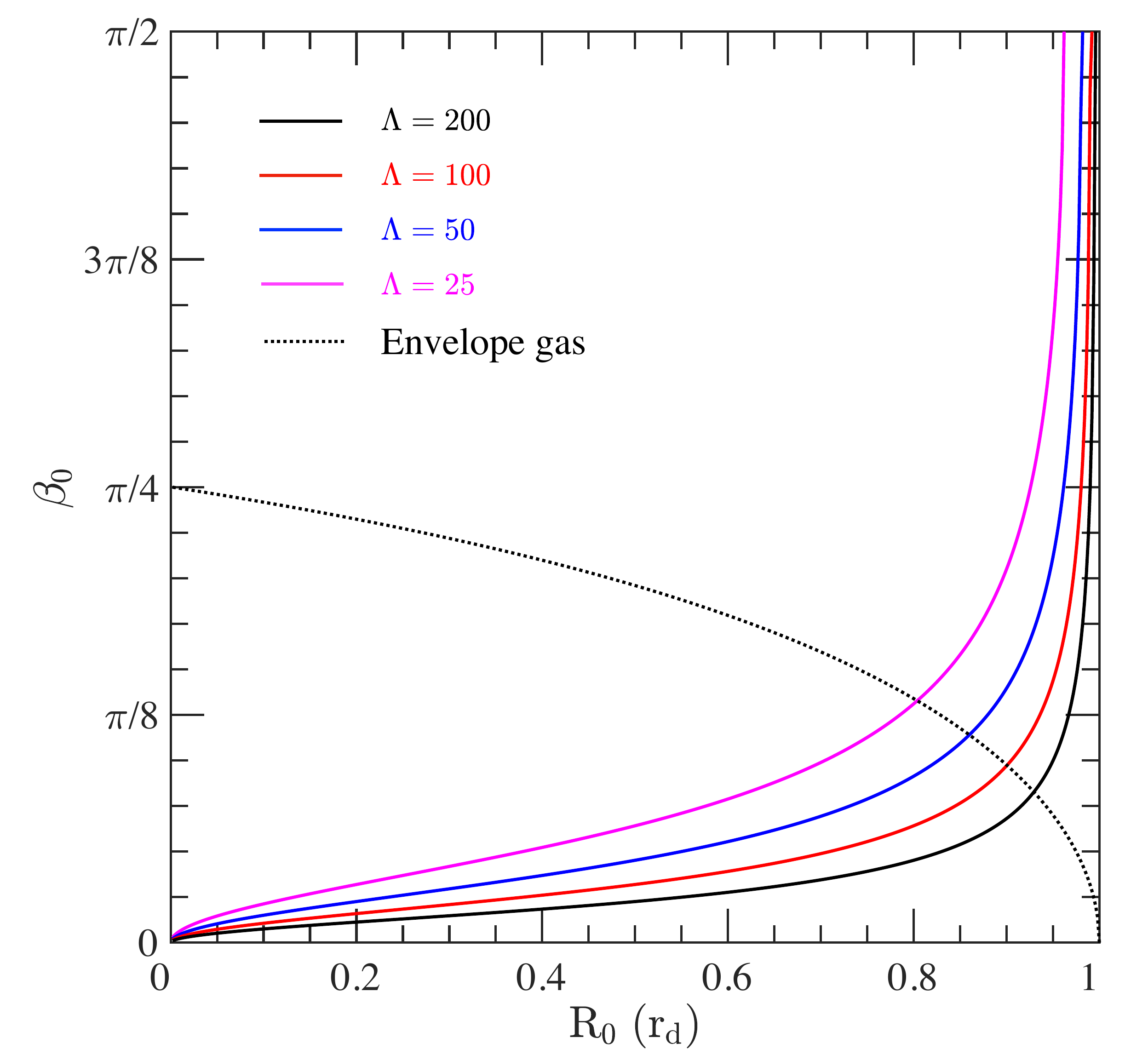}
    \vspace{-10 pt}
	\end{center}
    \caption{The allowed angle of incidence of the wind-envelope interface at the disk plane as a function of $R_0$, for models with different $\Lambda$ values. The dotted black line indicates the angle of incidence of the infalling material at $R_0$.}
    \label{fig:angle}
\end{figure*}

We determine the boundary condition of the interface at the disk plane by requiring pressure balance between the stellar wind and envelope sides. This directly leads to a one-to-one relation between $R_0$ (the distance of the wind/envelope interface from the central star at the disk plane) and $\beta_0$ (the angle of incidence at the base). Formally, 
\begin{equation}
\;\sin^2\beta_0 = \Lambda^{-1}\,(R_0/r_{\rm d})^{3/2}/(1-R_0/r_{\rm d}),
\end{equation}
which depends on $\Lambda$. In Figure~\ref{fig:angle} we show $\beta_0$ as a function of $R_0$ for $\Lambda=25$ (magenta line), 50 (blue line), 100 (red line) and 200 (solid black line). For purpose of reference, we also show in the same figure the local angle of incidence of the infalling envelope gas at the mid-plane as a function of $R_0$ (dotted black line).

It can be seen that for a given $\Lambda$, the allowed angle of incidence of the interface increases monotonically with $R_0$ and reaches $\pi/2$ near $r_d$ for all $\Lambda$ values that we have considered in this paper ($\beta_0=\pi/2$ indicates that the interface is  perpendicular to the disk plane). In the inner region, we find that the incidence angle of the envelope material at the disk plane becomes larger than the allowed $\beta_0$ of the wind/envelope interface. This corresponds to an unphysical solution where the envelope gas pushes the interface from the same side as the stellar wind. It is therefore a requirement that $R_0$ is sufficiently large so that a stable wind/envelope interface that is balanced by the pressure of the stellar wind and infalling envelope from either side is possible.

\begin{figure*}[ht]
    \begin{center}
    \includegraphics[width=0.6 \columnwidth]{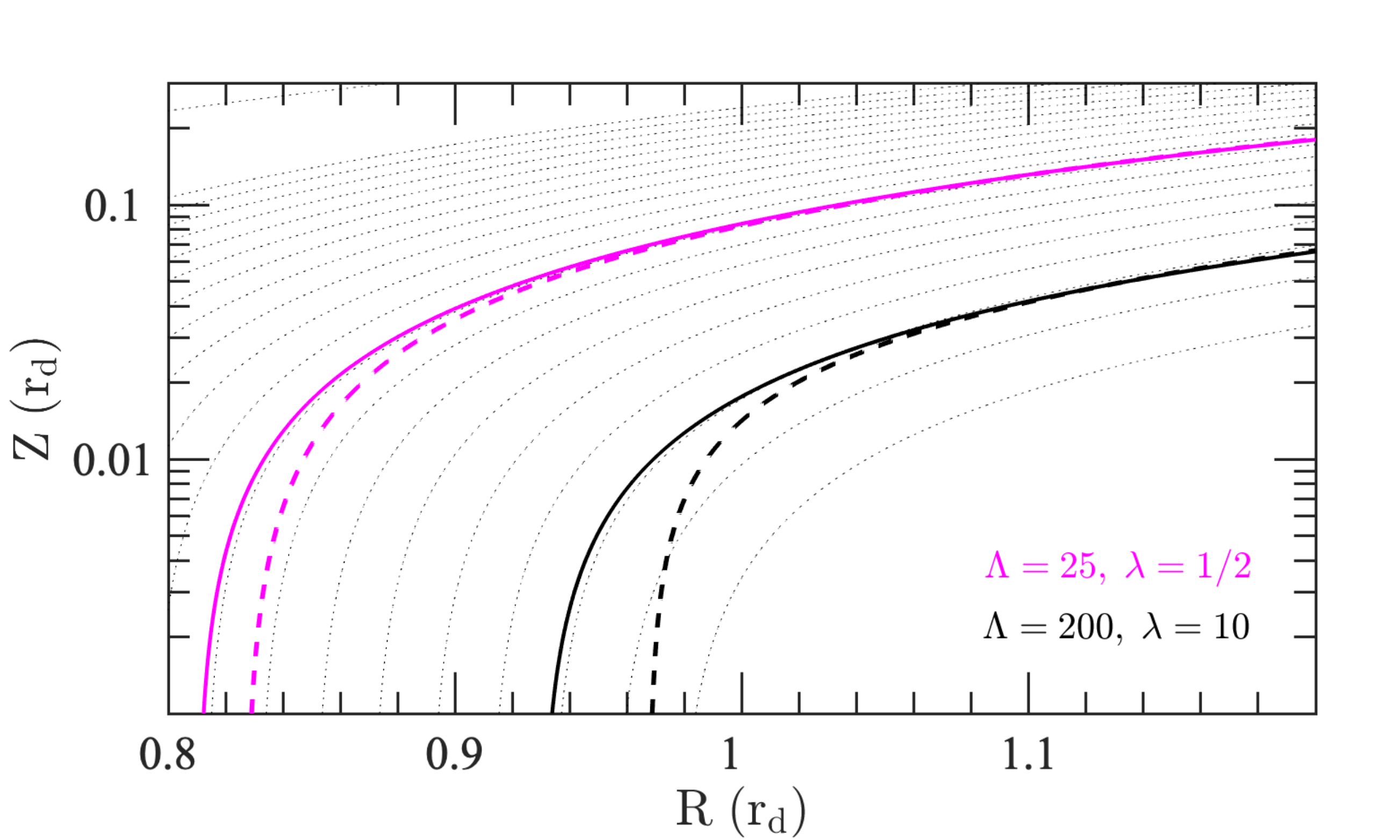}
    \vspace{-10 pt}
	\end{center}
    \caption{Cavity shapes near the base for an  isotropic  wind  colliding  with  an  \citet{ulrich76} infalling envelope for different $\Lambda$ values. The magenta and black lines correspond to the model of ($\Lambda=25$, $\lambda=1/2$) and ($\Lambda=200$, $\lambda=10$), respectively. For each model, the solid curve represents the `fiducial' solution where the wind/envelope interface at the foot-point lies parallel to the local streamline of the material of the infalling envelope at the disk plane, whereas the dashed line represents a different solution where the base of the interface locates at a larger $R_0$. For a given $\Lambda$, the solutions of different base position converge at a small distance from the disk plane (a few $0.01r_{\rm d}$). The thin dotted lines in the background indicate the streamlines of the material within the infalling envelope. }
    \label{fig:baseA}
\end{figure*}

For the fiducial models presented in the paper, the wind/envelope interfaces at the foot-point are parallel to the local streamline of the material of the infalling envelope at the disk plane (corresponding to the intersecting point between the dashed line and each colored solid line in Figure~\ref{fig:angle}). For a given $\Lambda$, when the interface foot-point is placed at somewhat larger $R$, the cavity quickly converges to the fiducial case within a small distance from the base, as is shown in Figure~\ref{fig:baseA}. The foot-point, however, cannot become arbitrarily close to $r_d$ without the infalling material crushing the wind and preventing a breakout solution, dependent on the value of $\lambda$. Thus the foot-point location is highly constrained, with larger $\lambda$ values allowing a broader range of solutions at the base, all converging to self-similar solutions at altitude. 

\section{Equations for the growth of the mixing-layer}
\label{ap:mixequ}

Several typographical errors were present in the general equations from \citet{rcc95} describing the growth of the mixing-layer between two axisymmetric moving fluids of speed $v_1$ and $v_2 < v_1$: in their equation (13), the term $hP^\prime$ should have been $-hP^\prime$, while in their equation (16), the factor $r_c$ next to $(h_1+h_2)P^\prime$ should not be present. Below, we reproduce their equations (15) and (16) where the latter typo has been corrected, and we use the subscript ``L" to denote quantities in the mixing-layer, instead of the lower case letter ``l", which was difficult to differentiate from the digit``1" in \citet{rcc95}. Furthermore, for consistency with the notation in the main paper, here we refer to the velocity within the mixing-layer as $v_L$ whereas in \citet{rcc95} it is just $v$. All other notations are kept the same. Since we assume an isothermal mixing-layer with uniform sound-speed $c_L$, we do not have to integrate their energy equation. Thus the system reduces to solving the following set of coupled equations for $h_1(x)$ and $h_2(x)$, which are the respective widths by which the mixing-layer encroaches into each fluid:
\begin{equation}
(\rho_1 v_1 - \rho_L \overline{v_L}) \frac{dh_1}{dx} + (\rho_2 v_2 - \rho_L \overline{v_L})\frac{dh_2}{dx} - (h_1+h_2)\overline{v_L} \frac{d\rho_L}{dx} = 
\rho_L (h_1+h_2) \frac{(\overline{v_L} r_c)^\prime}{r_c} - \alpha \rho_L c_L 
\end{equation}
and 
\begin{equation}
(\rho_1 v_1^2 - \rho_L \overline{v_L^2}) \frac{dh_1}{dx} + (\rho_2 v_2^2 - \rho_L \overline{v_L^2})\frac{dh_2}{dx} - (h_1+h_2)\overline{v_L^2} \frac{d\rho_L}{dx} = 
\rho_L (h_1+h_2) \frac{(\overline{v_L^2} r_c)^\prime}{r_c} - \alpha \rho_L c_L v_2
+ (h_1+h_2) P^\prime.
\end{equation}

In these equations, the subscript ``1" denotes quantities pertaining to the fast fluid (in our case, the deflected wind), the subscript ``2" pertains to the slow fluid (in our case, the deflected envelope), 
$x$ is the distance along the flow, $r_c$ and $P$ are the cylindrical radius and the pressure at the current point, primes denote derivatives versus $x$, and the mean velocities in the mixing-layer for a linear Couette flow are given by
\begin{equation}
\overline{v_L}  = \frac{v_1+v_2}{2}
\end{equation}
and
\begin{equation}
\overline{v_L^2}  = \frac{v_1^3-v_2^3}{3(v_1-v_2)}.
\end{equation}

The mass densities in the three layers are determined through transverse pressure equilibrium as
\begin{equation}
\rho_1 c_1^2 = \rho_2 c_2^2 = \rho_L c_L^2 = P,
\end{equation}
where we assume here for simplicity that the sound speeds $c_1$, $c_2$ and $c_L$ do not vary with position.

An added complication for our paper is that our two fluids do not flow in the same direction. Fortunately, we always have $v_2 < v_d  \ll v_1 \simeq v_{\rm w}$. We thus assume $v_2 = 0$ when integrating these equations upward along $x$, ie. that mass entrainment into the mixing-layer from the slow envelope side is largely dominated by the turbulent entrainment term. 


\section{Viscous dissipation in the mixing-layer}
\label{ap:mixheat}

We consider the simplified case, relevant to the present paper, 
of an isothermal, supersonic mixing-layer of width $h$ between two fluids with $v_2 \simeq 0$ and $v_1 \gg v_2$, and with a linear velocity gradient across the flow direction (Couette profile) $dv(y)/dy \simeq ({v_1}/{h})$ where $v_1$ changes weakly with position $x$ along the flow.  

We calculate $\Gamma_{\rm visc}$, the excess kinetic energy that needs to be locally dissipated by viscous turbulence per unit time and volume within the layer to maintain its internal linear Couette profile, as follows.
The flux of kinetic energy flowing along the mixing-layer is given by
\begin{equation}
\begin{split}
   \dot E (x) & = 2 \pi R \int_0^h {\rho_L \left[ \frac{v(y)^3}{2} \right] dy },\\
   & = 2 \pi R \rho_L h \frac{v_1}{2} \left[\frac{v_1^2}{4}\right],\\
   & = \dot M_L(x) \left[\frac{v_1^2}{4} \right].
\label{eqn:kinflux}
\end{split}
\end{equation}
Note that in converting to units of mass flow in the layer, $\dot M_L$, we make use of the fact that the mean flow velocity in the layer is $\overline{v_L} = v_1/2$ (Couette profile
with $v_2 \simeq 0$).

We next consider a "slice" of mixing-layer of thickness $\Delta x$, and denote
$\dot E_{\rm in}$ and $\dot E_{\rm out}$ the kinetic energy flux flowing through the layer at positions $x$ and $x+\Delta x$, respectively. Since $v_1$ is considered constant with position, 
the change in kinetic energy flux (Eqn.~\ref{eqn:kinflux}) between $x$ and $x+\Delta x$ is 
caused only by the increase in mass-flux through the layer, $\Delta \dot M_L$, via sideways entrainment. Recalling that, to maintain a Couette flow with $v_2 \simeq 0$, the entrainment rate from the wind side must be twice that from the ambient side (see Section~\ref{sec:flow:mix}), we then have 
\begin{equation}
\begin{split}
   \dot E_{\rm out} - \dot E_{\rm in} & = \Delta \dot M_L \left[ \frac{v_1^2}{4}  \right] = 3 \rho_2 v_{\rm ent}  (2 \pi R \Delta x) \left[\frac{v_1^2}{4} \right]. 
\label{eqn:delkinflux}
\end{split}
\end{equation}
At the same time, the kinetic energy flux $\dot E_{\rm ent}$ injected into the slice through its lateral surfaces by entrainment of fresh wind material at $v_1$ is given by 
\begin{equation}
\begin{split}
   \dot E_{\rm ent}  & = 2 \rho_2 v_{\rm ent}  (2 \pi R \Delta x) \left[\frac{v_1^2}{2} \right].
\end{split}
\label{eqn:delkinent}
\end{equation}
The excess injected kinetic energy that needs to be dissipated by turbulent viscosity per unit time within the slice volume to maintain the Couette flow is 
\begin{equation}
\begin{split}
   \dot E_{\rm visc} & =  \dot E_{\rm ent} + \dot E_{\rm in} - \dot E_{\rm out}
    \equiv \Gamma_{\rm visc} (2 \pi R \Delta x) h. 
\end{split}
\label{eqn:delkindiss}
\end{equation}
Combining Equations \ref{eqn:delkinflux}, \ref{eqn:delkinent} and \ref{eqn:delkindiss}, we obtain 
\begin{equation}
\begin{split}
   \Gamma_{\rm visc}  & =   \frac{\rho_2 v_{\rm ent}}{h} \left[v_1^2 - \frac{3v_1^2}{4} \right]\\
                 & = \frac{\rho_2 v_{\rm ent}}{h} \left[\frac{v_1^2}{4} \right]
                 = \frac{\alpha}{4} \rho_L c_L {h} \left(\frac{v_1}{h} \right)^2, 
\end{split}
\label{eqn:gammavisc} 
\end{equation}
where we make use of our prescription for $v_{\rm ent} \equiv \alpha\, c_2^2 / c_L$ and recognize the lateral pressure equilibrium across the shell, $\rho_2 c_2^2 = \rho_L c_L^2$. 
Comparing Eqn.~\ref{eqn:gammavisc} with the standard expression for viscous dissipation,
$\Gamma_{\rm visc} = \mu \left({dv(y)}/{dy} \right)^2$, we obtain an ``effective" turbulent viscosity
in the mixing-layer 
\begin{equation}
\mu = \frac{\alpha}{4} \rho_L c_L {h}.
\label{eqn:viscosity} 
\end{equation}
Note that we recover the same turbulent viscosity prescription as in Equation (5) of the work of \citet{Binette99} (with their parameter $\alpha$ being 1/4th of our $\alpha$). 
For the typical $\alpha = 0.03$ favored by mixing-layer experiments \citep{cr91}, the numerical coefficient in Eqn.~\ref{eqn:viscosity} would be $\simeq 0.007$, as adopted by \citet{Binette99} for their calculations. 
\end{document}